\documentclass[epj]{svjour_mod}
\usepackage{geometry}                
\usepackage{graphicx}
\usepackage{amssymb}
\usepackage{stfloats}
\usepackage{xspace}
\usepackage{url}
\usepackage{cite}
\usepackage{float}
\usepackage{multirow}
\usepackage{booktabs}
\usepackage{graphics}
\usepackage{epstopdf}

\begin{document}
\include{00README.XXX}
\title{A VUV detection system for the direct photonic identification of the first excited isomeric state of $^{229}$Th}
\subtitle{}
\author{Benedict Seiferle\inst{1}\thanks{\emph{Present address:} benedict.seiferle@physik.uni-muenchen.de} \and Lars von der Wense\inst{1} \and Mustapha Laatiaoui\inst{2,}\inst{3} \and Peter G. Thirolf\inst{1}
}                     
%
%
\institute{Ludwig-Maximilians-Universit\"at M\"unchen, Am Coulombwall 1, Garching, Germany \and %
              GSI Helmholtzzentrum f\"ur Schwerionenforschung GmbH, Planckstr. 1, Darmstadt, Germany \and %
              Helmholtz Institut Mainz, Johann-Joachim-Becherweg 36, Mainz, Germany}
\date{November 24, 2015}
%
\authorrunning{B. Seiferle \textit{et al.}}
\titlerunning{A VUV detection system for the direct identification of  $^{229m}$Th}
\abstract{%
With an expected energy of 7.6(5) eV, $^{229}$Th possesses the lowest excited nuclear state in the landscape of all presently known nuclei. 
The energy corresponds to a wavelength of about 160 nm and would conceptually allow for an optical laser excitation of a nuclear transition.
We report on a VUV optical detection system that was designed for the direct detection of the isomeric ground-state transition of $^{229}$Th. 
$^{229(m)}$Th ions originating from a $^{233}$U $\alpha$-recoil source are collected on a micro electrode that is placed in the focus of an annular parabolic mirror. 
The latter is used to parallelize the UV fluorescence that may emerge from the isomeric ground-state transition of $^{229}$Th. 
The parallelized light is then focused by a second annular parabolic mirror onto a CsI-coated position-sensitive MCP detector behind the mirror exit.
To achieve a high signal-to-background ratio, a small spot size on the MCP detector needs to be achieved. 
Besides extensive ray-tracing simulations of the optical setup, we present a procedure for its alignment, as well as test measurements using a D$_2$ lamp, where a focal-spot size of $\approx$100 $\mu$m has been achieved. %
Assuming a purely photonic decay, a signal-to-background ratio of $\approx$7000:1 could be achieved.
\PACS{ {PACS-key} 
      			{23.35.+g  Isomer decay	
			}   \and {}%
     			{27.90.+b 	A $>$ 220
			}   \and {}%
     			{42.15.Eq 	Optical system design
			}  \and {}
			{06.60.Sx	Positioning and alignment; manipulating, remote handling}
} 
} 
\maketitle 
\section{Introduction} \label{Introduction}
In 1976, Kroger and Reich proposed the existence of an isomeric state in $^{229}$Th with an energy below 100 eV \cite{PropKrog}.
Since then, the knowledge on the ground-state transition energy of this isomeric state based on indirect measurements (\textit{i.e.} the comparison of $\gamma$ lines of higher energy) has been gradually extended.
In 1990, such a measurement revealed a value of 1$\pm$4 eV \cite{1990}. 
Benefitting from higher statistics, this value was refined to 3.5$\pm$1.0 eV \cite{1994} in 1994, which corresponds to a wavelength of about 350 nm. 
In 2007, another indirect measurement with an improved detector energy-resolution was performed and placed the ground-state transition energy at 7.6$\pm$0.5 eV \cite{Beck7.6eV} (updated to 7.8 eV in \cite{7.8eV}). 
This energy value corresponds to a wavelength of about 160 nm and is the accepted value until today.
Despite numerous attempts \cite{Irwin, Richardson, Utter, Shaw, 2001, 2003, 2003_1, 2003_3, 2004, 2005, 2005_2, Zimmermann, Swanberg, 6+-1, PeikComment}, however, %
there has not been an undoubted direct energy measurement of the isomeric ground-state transition of $^{229} $Th up to today. 
This is partly due to the fact that before 2007 the wavelength region around 350 nm was investigated (corresponding to 3.5 eV). 
From today's point of view these experiments could not succeed, since the energy of 7.6 eV shifts the wavelength into the vacuum ultra-violet (VUV) region, that requires appropriate optical elements.\\
With its expected unique properties ($\Delta$E=7.6 eV and $\tau\approx 10^4$ s \cite{Lifetime}), $^{229(m)}$Th could conceptually link nuclear physics with methods and applications known from atomic physics, such as laser spectroscopy and quantum metrology. 
Due to its long lifetime, $^{229(m)}$Th exhibits an exceptionally narrow line-width $\Delta E/E$ in the order of $10^{-20}$.
This narrow line-width, and also the possibility to drive the transition with lasers, makes $^{229(m)}$Th an interesting candidate for a nuclear optical clock \cite{Peik1, Kazakov, Campbell}.
The latter could be used, for example, to measure potential temporal variations of fundamental constants with improved sensitivity \cite{Flammbaum, Flammbaum1, Flammbaum2}, by comparing them with atomic optical clocks that use electronic shell transitions.
In case of the fine structure constant $\alpha$, the sensitivity is presently limited to $\delta \alpha/\alpha=10^{-17}/\mbox{a}$ when using transitions in the electronic shell \cite{Rosenbund}. The sensitivity could be improved to $\delta \alpha/\alpha<10^{-20}/\mbox{a}$, when the isomeric ground-state transition of $^{229}$Th was used \cite{Flammbaum}.
There are also thoughts towards a nuclear $\gamma$-ray laser in the optical range that uses the isomeric transition of $^{229} $Th\cite{Tkalya-gamma}.
Still, progress on the way towards these applications is detained by the relatively vague knowledge on the isomer's energy, that needs to be known to better than about 0.1 eV in precision in order to develop a laser that may drive the extremely narrow transition.
\\The long-term goal of the experiment, which is subject of this paper, is the detection of the expected VUV radiation that is emitted by the isomeric ground-state transition and the determination of its wavelength. Therefore, an optical setup that consists of two annular parabolic mirrors has been developed, extensively tested and characterized. 
In the following, an overview of the experimental setup, a study of the influence of its possible misalignment, the alignment procedure, as well as measurements to verify the properties of the optical setup are presented. These measurements are leading to an unsurpassed conceptual signal-to-background ratio of 7000:1.
\section{Experimental setup}
	\begin{figure*}[ht]
	\centering
	\includegraphics[width=0.9 \textwidth]{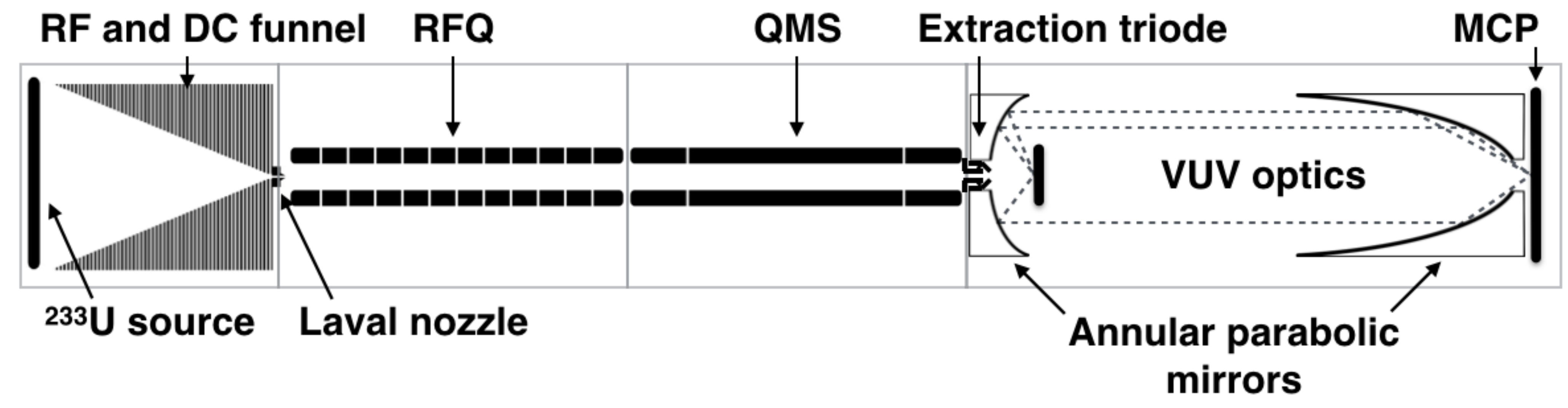}
	\caption{\label{setupExtraction} Overview of the experimental setup. The first chamber ($p=40$ mbar) contains the $^{233}$U source and the RF and DC funnel. A supersonic laval nozzle marks the entrance to the second chamber ($p\approx10^{-2}$ mbar) that contains the segmented RFQ. The next chamber ($p\approx10^{-5}$ mbar) houses a quadrupole mass separator (QMS). 
	The components in the last chamber which is the optics chamber ($p\approx10^{-6}$ mbar) are detailed in fig. \ref{SketchOptics}. }
	\end{figure*}
\noindent
The conceptual idea of the setup is to first extract $^{229(m)}$Th from a $^{233}$U $\alpha$-recoil ion source and form an ion beam out of the recoil ions. $^{229m}$Th is populated by a 2\% decay branch of the $^{233}$U $\alpha$-decay \cite{Barci}. The $^{229(m)}$Th ions are then collected on a $\O$=50 $\mu$m micro electrode, that is placed in the focus of an annular parabolic mirror, which is part of an optical system. This optical system focuses the expected isomeric ground-state decay radiation on a VUV- and position-sensitive multi-channel plate (MCP) detector. 
\subsection{Extraction of $^{229(m)}$Th from a $^{233}$U $\alpha$-recoil source}\label{Extraction}
The components for the formation of a $^{229(m)}$Th ion beam are housed in three differentially pumped vacuum chambers, namely a buffer-gas stopping cell (40 mbar He), a radio-frequency quadrupole (RFQ) and a quadrupole mass separator (QMS) chamber (operated at $\approx10^{-2}$ mbar and $\approx10^{-5}$ mbar, respectively). Fig. \ref{setupExtraction} shows a scheme of the experimental setup:
A $^{233}$U $\alpha$-recoil ion source (290 kBq) is placed inside the buffer-gas stopping cell, that is filled with 40 mbar ultra pure helium (6.0) that stops the ions. 
The buffer-gas stopping cell is built to fulfill UHV standards and is bakeable up to 180$^\circ \mbox{C}$. The helium gas is catalytically purified to the ppb level and can be further purified by making use of a cryotrap and a getter pump, before it is injected into the cell via electropolished tubes.
Radio frequency (220 V$_{\mbox{pp}}$ at 850 kHz) and DC voltages are applied to 50 ring-electrodes that are arranged in a funnel-like structure (RF and DC funnel) to guide the ions towards a supersonic Laval-nozzle that links the buffer-gas stopping cell with the next chamber. There, a continuous ion beam is produced by a segmented radio frequency quadrupole ion-guide (RFQ), operated with 200 V$_{\mbox{pp}}$ at 880 kHz. 
Different DC voltages applied to each segment create a gradient (-0.2 V per segment) that drags the ions through the remaining phase-space cooling buffer gas.
Next, the ion beam is mass selected by a quadrupole mass separator (QMS) which was built according to the design values reported in ref. \cite{Haettner}. In this way, no other daughter isotopes from the $^{233}$U decay-chain than $^{229(m)}$Th reach a triodic extraction nozzle that marks the entrance to the optics chamber. 
The triodic extraction nozzle exhibits a compact geometry of 10 mm in diameter for all electrodes.
In this way it fits through a 12 mm central hole of an annular parabolic mirror (see next section and fig. \ref{SketchOptics}), in order to achieve a short distance of 3 mm between the extraction nozzle and a \O\xspace=50 $\mu$m micro electrode on which the ions are collected.
A more detailed explanation of the extraction part can be found in ref. \cite{Lars}, which reports the combined extraction and purification efficiency for $^{229} $Th in the 3+ charge state to be 10 \%.
%
%
\subsection{VUV optics}\label{VUV optics}
\begin{figure*}[ht]
	\centering
	\resizebox{0.9\textwidth}{!}{%
	\includegraphics{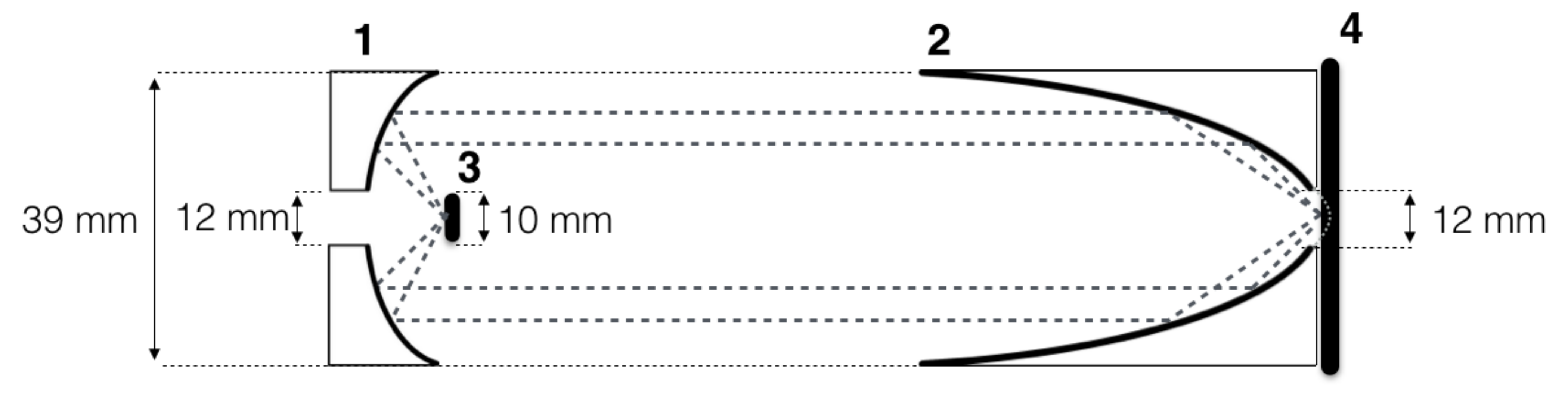}}
	\caption{Sketch of the VUV optical setup. It consists of two annular parabolic mirrors (\textbf{1} and \textbf{2}), a collection surface (\textbf{3}) with an onprinted 50 $\mu$m micro electrode and a multi-channel plate (MCP) detector (\textbf{4}). The first (shallow) annular parabolic mirror (\textbf{1}) parallelizes the light that may originate from the decay of $^{229m}$Th, which was collected on the micro electrode. The second (deep) parabolic mirror (\textbf{2}) focuses the collimated light onto a MCP detector (\textbf{4}). The beam path is indicated by the dashed lines. All measures are given in mm.\label{SketchOptics}}
	\end{figure*}

\begin{figure}[H]

	\begin{minipage}[ht]{0.48\textwidth}
	\centering
	\includegraphics[width=1\textwidth]{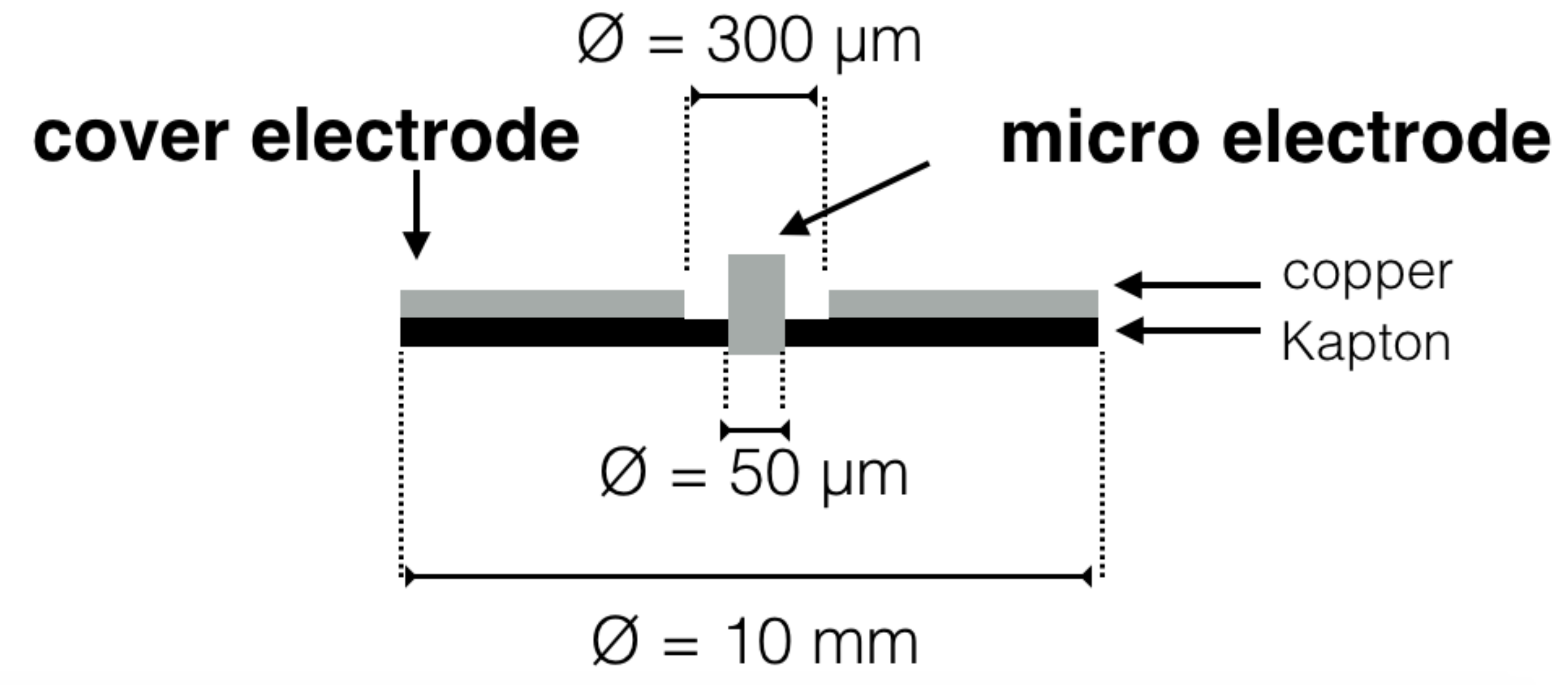}
	\end{minipage}
	\caption{\label{CollElecMac} Profile of the collection surface. Elements made of Kapton are colored in black, copper is indicated in grey. The thickness of the copper layer is $\approx$30 $\mu$m.}
	\end{figure}

\begin{figure*}[ht]
	\begin{minipage}[ht]{0.24\textwidth}
	\begin{flushleft} \textbf{  a} \end{flushleft}
	\centering
	\includegraphics[height=0.9\textwidth]{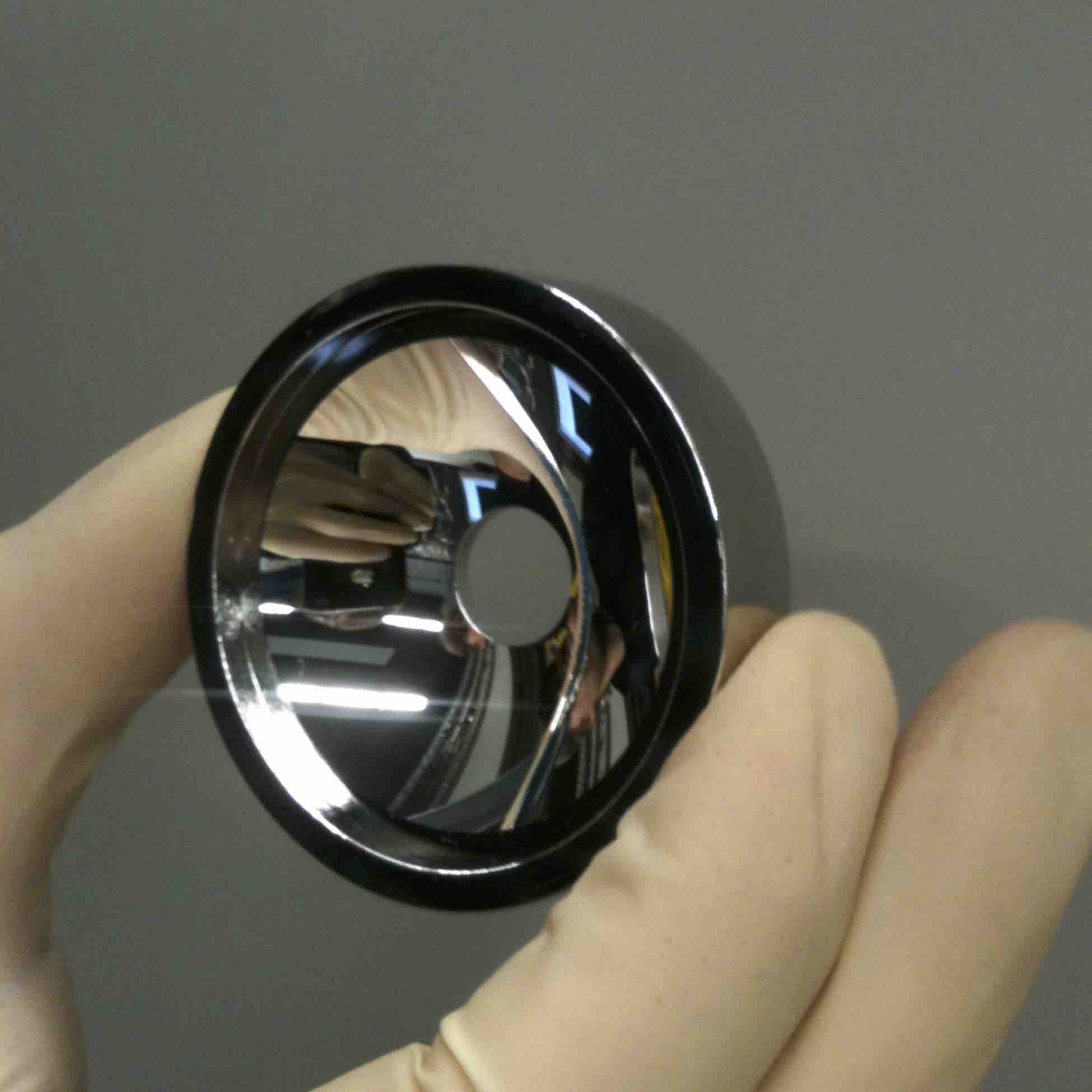}
	\end{minipage}
	\begin{minipage}[ht]{0.24\textwidth}
	\begin{flushleft} \textbf{b} \end{flushleft}
	\centering
	\includegraphics[height=0.9\textwidth]{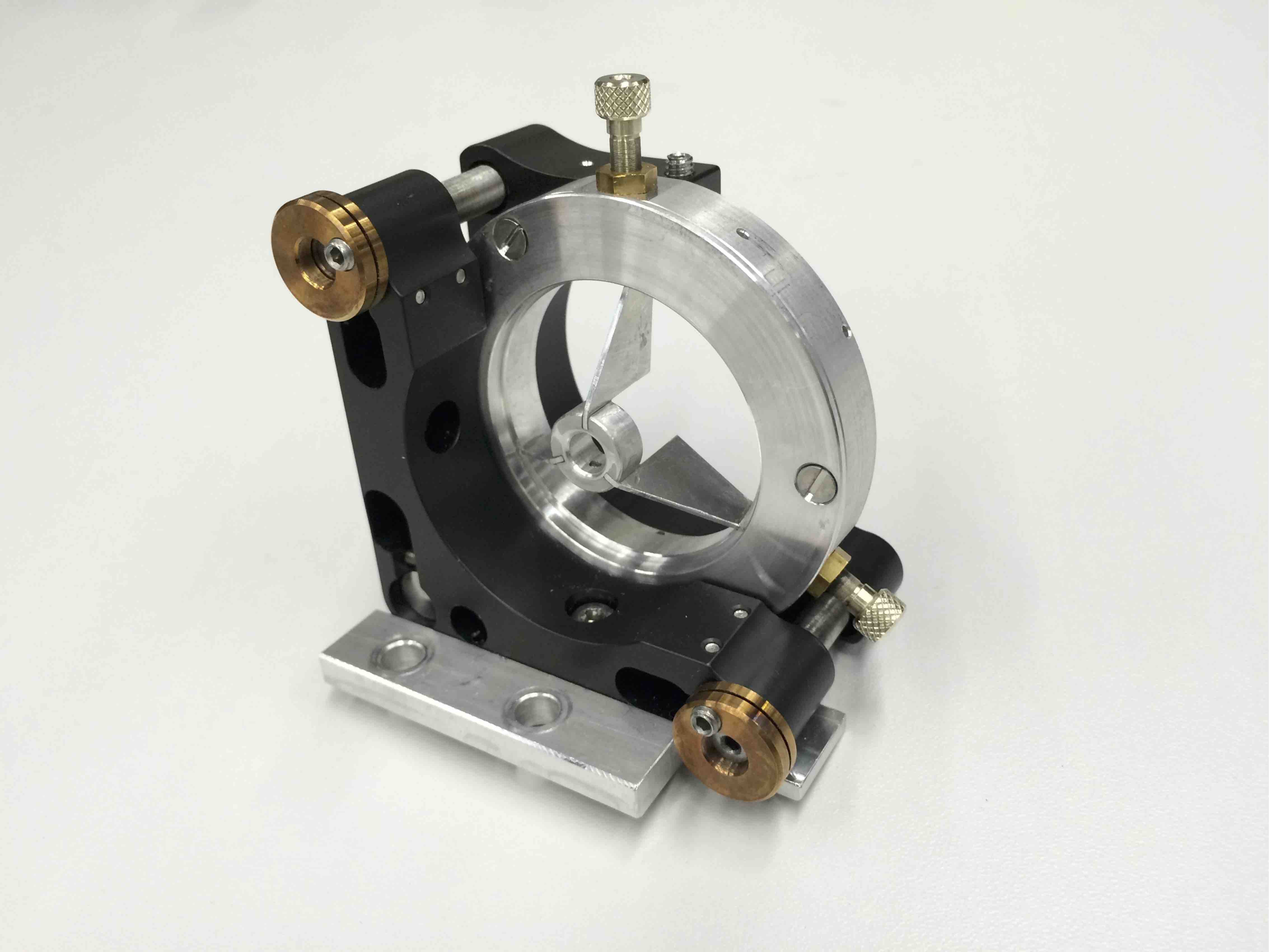}
	\end{minipage}
	\begin{minipage}[ht]{0.24\textwidth}
	\begin{flushleft} \textbf{  c} \end{flushleft}
	\centering
	\includegraphics[height=0.9\textwidth]{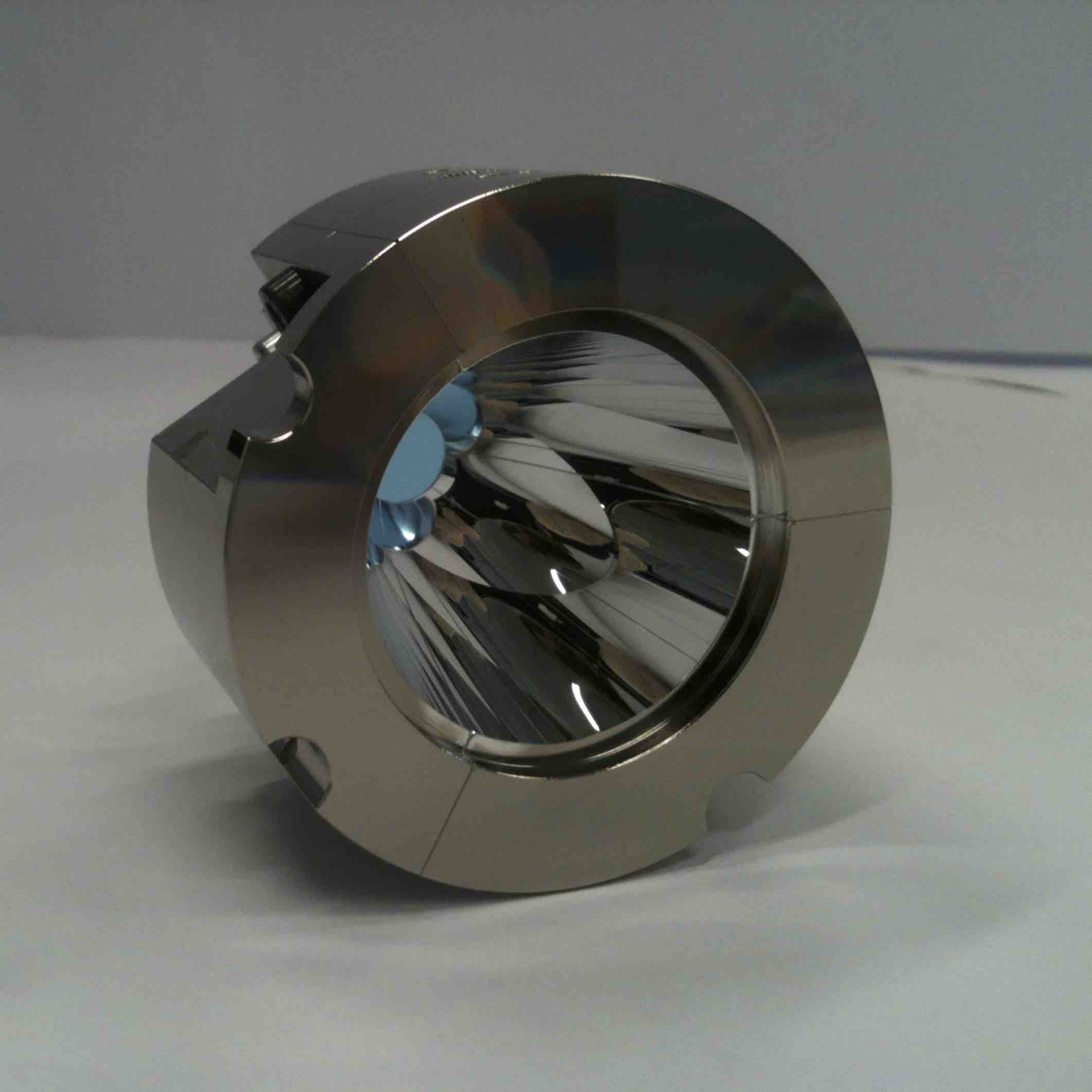}
	\end{minipage}
	\begin{minipage}[ht]{0.24\textwidth}
	\begin{flushleft} \textbf{d} \end{flushleft}
	\centering
	\includegraphics[height=0.9\textwidth]{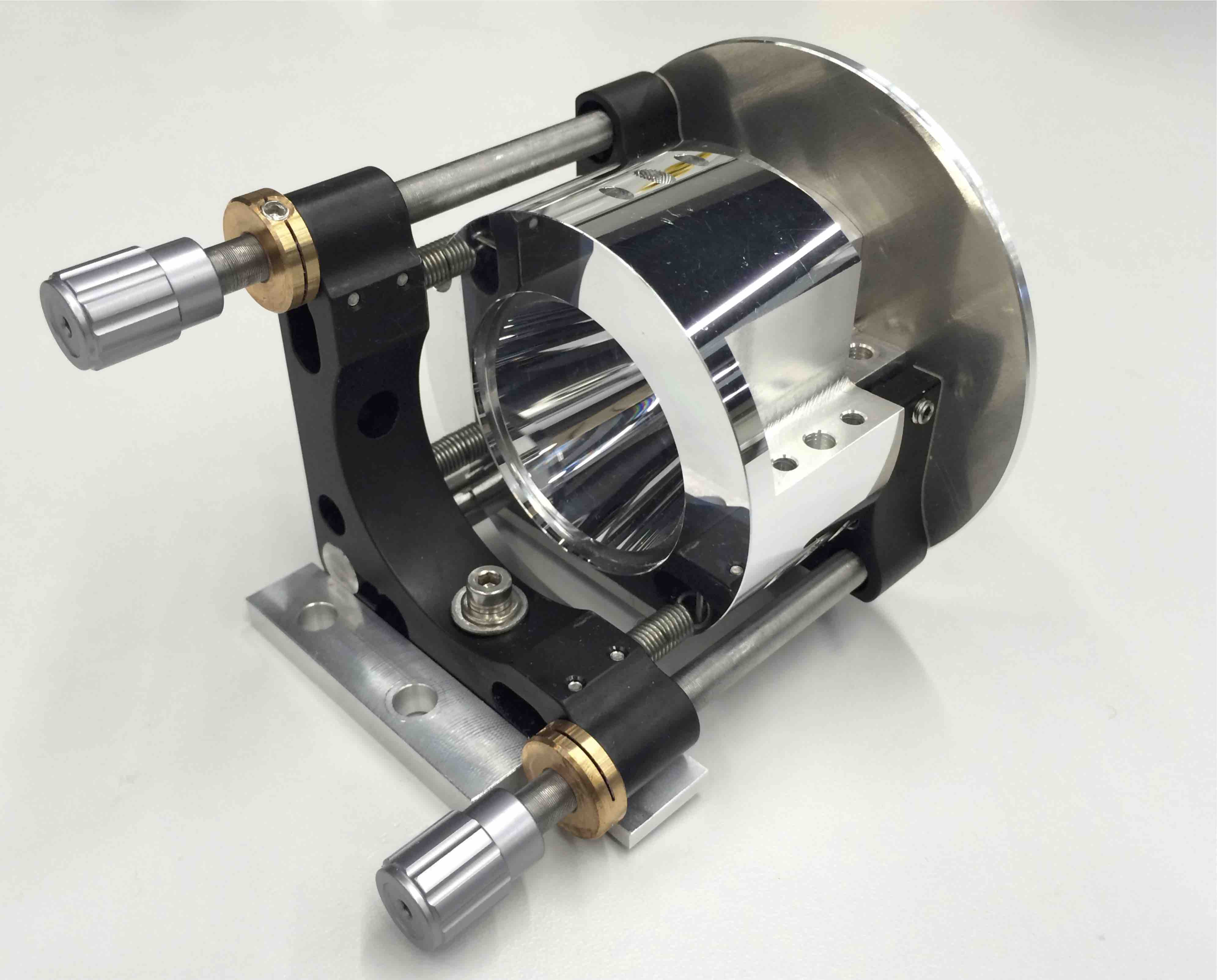}
	\end{minipage}
	\caption{\label{Mirrors} Picture of the VUV mirrors and their holders. \textbf{a}: shallow annular parabolic mirror (mirror 1). \textbf{b}: mirror mount of mirror 1 with the tripod that serves as a holder for the collection surface. \textbf{c}: deep annular parabolic mirror (mirror 2). \textbf{d}: mount of mirror 2. }
	\end{figure*}

%
An optical setup for the direct detection of the isomeric ground-state transition in $^{229}$Th needs to fulfill several requirements.\\
Firstly, with the 2\% decay branch from the $^{233}$U $\alpha$ decay, one expects a rather low isomeric ground-state decay rate. 
In addition, there is no experimental knowledge on the internal conversion (IC) and the bound internal conversion (BIC) coefficients of the transition ($\alpha_{IC}=\Gamma_{IC}/\Gamma_{\gamma}$ and $\alpha_{BIC}=\Gamma_{BIC}/\Gamma_{\gamma}$, respectively). For a ground-state transition energy of 7.6 eV and neutral $^{229}$Th atoms, internal conversion is expected to be the dominant decay channel with $\alpha_{IC}\approx10^9$ \cite{Karpeshin}. Therefore it is envisaged to suppress the IC channel by collecting the ions on a MgF$_2$-coated electrode surface. 
In this way, although collected on the electrode surface, $^{229}$Th should remain in a charged state when it is stopped on the crystal surface. This is expected to hinder the IC decay channel, as the second ionization potential of $^{229}$Th (given by 11.9 eV \cite{SEC-Ionization}) is greater than the expected energy of the isomer. 
Due to the expected low count rates, it is of major importance to provide a high light yield onto the detector with the anticipated optical setup. To reach a high signal-to-background ratio, the photons are focused onto a small spot on a position-sensitive detector.\\
Secondly, the energy of the isomeric transition is not yet precisely known, and such is the wavelength of the light that is emitted by the de-excitation of the isomer.
For this reason, it is of great importance to design an optical setup which keeps optimized properties and efficiencies over a wide wavelength region. 
However, the light is expected to lie in the VUV region ($\lambda <$ 200 nm), which requires a vacuum ultra-violet optical system.\\
With these considerations in mind, the optical setup was designed as schematically shown in figs. \ref{setupExtraction} \& \ref{SketchOptics}.
\noindent
 $^{229(m)}$Th ions, extracted from the $^{233}$U source, are collected on a \O\xspace=50 $\mu$m micro electrode collection surface (\textbf{3}). 
The micro electrode is placed in the focal plane of a (shallow) annular parabolic mirror (mirror 1) (\textbf{1}). When the isomeric decay of $^{229m}$Th occurs on the micro electrode, it can be assumed as a nearly point-like light source. 
The light is then parallelized by mirror 1 and subsequently focused by a second annular parabolic mirror (mirror 2) (\textbf{2}) onto the detector (\textbf{4}). For this purpose, the detector is placed in the focal plane of the second mirror. \\
Such a setup, solely based on parabolic mirrors, does not suffer from chromatic or spherical aberrations. 
With the all-reflective surfaces of the mirrors (for MgF$_2$ coated mirrors the reflectivity is larger than 70 \% for $\lambda \gtrsim$130 nm), we are provided with an optical system whose properties are unaffected by the isomer's energy, if the corresponding energy is below $\approx$ 9.5 eV ($\hat{=}$ 130 nm).
The focal lengths of the mirrors are g=10 mm for the first and b=2 mm for the second mirror, respectively.
In the paraxial approximation, a magnification of $\mathcal{M}=b/g=2/10=0.2$ is calculated. However, since the realistic optics is highly non-paraxial, the paraxial approximation does not apply here and by ray tracing simulations (see sect. \ref{Ray-tracing}) a magnification of $\mathcal{M}=1.4$ is determined.
\\
The collection surface consists of a $\O$=10 mm  two-layer (copper and Kapton) printed circuit board. 
The micro electrode was machined by etching off a $\O$=300 $\mu$m ring of the copper layer, leaving a $\O$=50 $\mu$m spot in the center (\textit{i.e.} the micro electrode). 
In this way, the micro electrode is electrically separated from the rest of the collection surface, which itself acts as a shielding electrode (referred to as  ``cover electrode'' in the following sections).
Both electrodes can be connected from the backside.
A sketch of the profile of the collection surface is shown in fig. \ref{CollElecMac}. A microscopic picture of the micro electrode is shown later in the left panel of fig. \ref{collectionMicroscope} (in sect. \ref{AlignmentProc}). The collection surface is mounted to a vacuum stepper motor (\textit{Micromotion Micro Linear Pusher} \cite{Micromotion}), to externally control its position in the beam direction (0.25 $\mu$m smallest achievable translation). The motor is controlled with a LabView program and is attached to the mount of mirror 1 with a tripod that allows for a positioning relative to mirror 1 perpendicular to the beam direction. 
The tripod can be seen in fig. \ref{Mirrors} \textbf{b}. 
%
\noindent Mirror 1 covers 42\% of the solid angle seen from the micro electrode and thus ensures a high light yield.
Mirror 2 works as a focusing mirror: its shape follows a parabola, where the vertex was cut out in order to uncover the focal point, which lies 2 mm behind the mirror exit (see fig. \ref{SketchOptics}).
Both mirrors were manufactured and characterized by the \textit{Fraunhofer-Institut f\"ur Angewandte Optik und Feinmechanik} (IOF) \cite{Fraunhofer}. Their surfaces are made of MgF$_2$-coated aluminum (surface roughness of 1 nm (rms)) and have a high reflectivity ($>$70\%) for wavelengths larger than 130 nm \cite{Coating}. 
Both mirrors are 39 mm in diameter. 
The mirror's focal lengths are 10 mm and 2 mm for mirror 1 and mirror 2 respectively. There is a large difference in the shape deviations that account to 200 nm p.v. (peak to valley) for mirror 1 and 7.7 $\mu$m p.v. for mirror 2 which is due to the production process. The surface of mirror 2 is too steep to allow for an effective coating using a regular coating procedure. Therefore the mirror was cut into three segments, each segment was coated separately to be merged later on, leading to the relatively large, but still tolerable, shape deviations. Both mirrors are mounted to customized mirror mounts (\textit{NewPort U200-A2H ULTIMA Clear Quadrant Mirror Mount}), to meet the requirements that are determined by the special mirror geometries. The mirrors and their mountings are shown in fig. \ref{Mirrors}.
	\noindent
%
%
Each mirror is attached via its mounting to a positioning stage, that itself is mounted to an optical base plate. Both stages are motorized (\textit{Faulhaber AM 2224} with a \textit{BS22-1.5 4864} gearing for mirror 1 and \textit{pi-micos MP20B linear pusher} for mirror 2). In this way, mirror 1 and mirror 2 can be positioned in the beam direction with a precision of 60 $\mu$m and 0.54 $\mu$m, respectively. A LabView program was developed for the purpose of an external position control of the mirrors and the micro electrode.
The detector's position is fixed, therefore the second mirror needs to be moved, so that its focal spot falls onto the detection plane. 
The detector is easily replaceable and can be adapted to the isomer wavelength. In the initial setup, a VUV-sensitive, CsI-coated MCP detector with a phosphor screen and a channel diameter of 25 $\mu$m is used and monitored by a CCD camera. The typical quantum efficiency of CsI-coated MCP detectors is above 10\% for wavelengths below 160 nm. 
The arrangement of the optical system leaves space between the two mirrors for setups to determine the wavelength of the parallelized fluorescence radiation, like, \textit{e.g.}, filters with sharp absorption edges.
\\
\section{Alignment of the optical system} \label{Alignment of the optics}
The alignment of the optical system poses a special challenge, because first the optical axis is blocked by the micro electrode, second the optical setup is highly non-paraxial and third a part of the alignment has to be performed under vacuum. For these reasons a customized alignment procedure was developed.
We define the beam/optical axis as the z axis, which forms an orthonormal basis with the x- and y axis. The angle $\theta$ describes the tilting with respect to the z axis. $\varphi$ is the angle of rotations around the z axis. The coordinate system is illustrated in fig. \ref{AlignmentIllustrationA}.
The optical axis (z axis) concurs with the symmetry axis of the openings of the vacuum chamber that is housing the optical setup. 
The detector is assumed to be inherently aligned, since it is mounted directly onto the chamber exit, thus the optical axis is orthogonal to the detection plane. 
When aligning the optical system, one has to deal with six relevant degrees of freedom: $\theta$ of mirror 1 and mirror 2, the x, y and z position of the micro electrode with respect to mirror 1 and the z position of mirror 2 with respect to the detector. All remaining degrees of freedom do not require a high precision. These are the x-y positions of the mirrors, all degrees of freedom of the MCP detector, the z position of mirror 1 and the  $\theta$ orientation of the micro electrode. One needs to mention that the optical setup possesses rotational symmetry around the optical axis when the $\theta$-orientation of each component is aligned, such that misalignments in $\varphi$ will not make a difference and will therefore be ignored for the rest of the discussion.
\subsection{Ray tracing simulations and influence of possible misalignments}\label{Ray-tracing}
\begin{figure*}[ht]
	\begin{minipage}[ht]{0.33\textwidth}
	\flushleft\textbf{~~~~a}
	\includegraphics[width=1\textwidth]{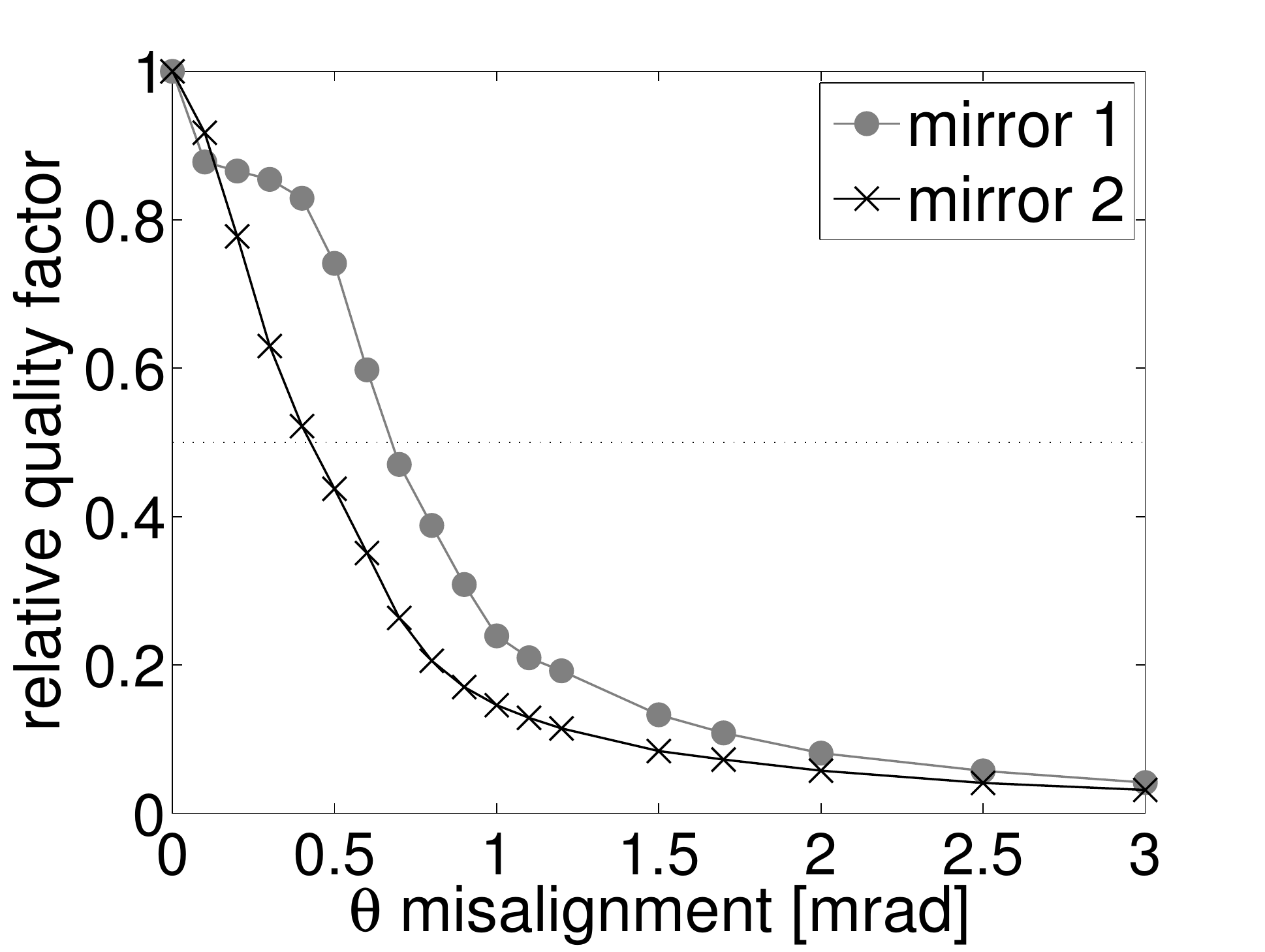}
	\end{minipage}
	\begin{minipage}[ht]{0.33\textwidth}
	\flushleft\textbf{~~~b}
	\includegraphics[width=1\textwidth]{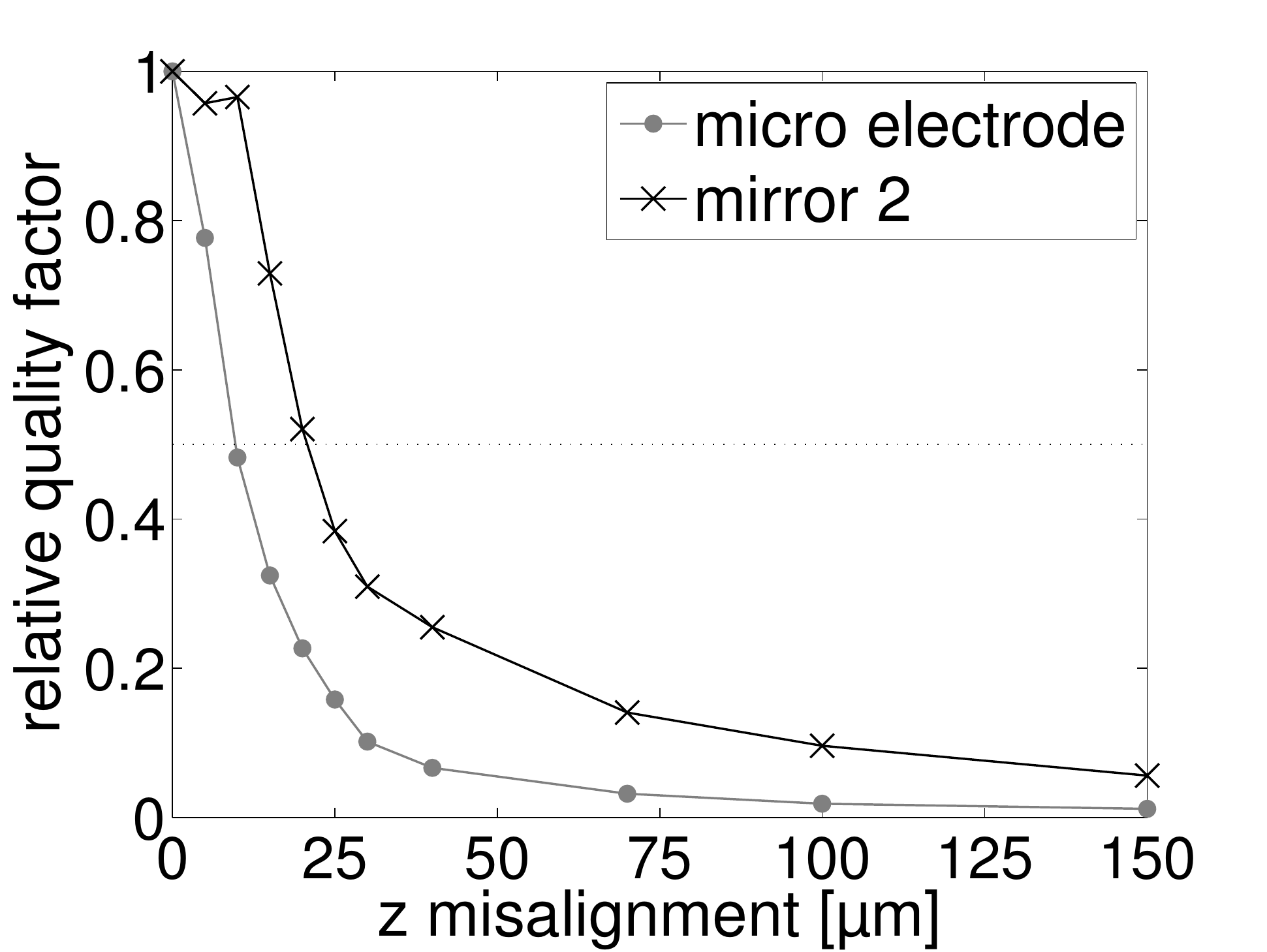}
	\end{minipage}
	\begin{minipage}[ht]{0.33\textwidth}
	\flushleft\textbf{~~~~c}
	\includegraphics[width=1\textwidth]{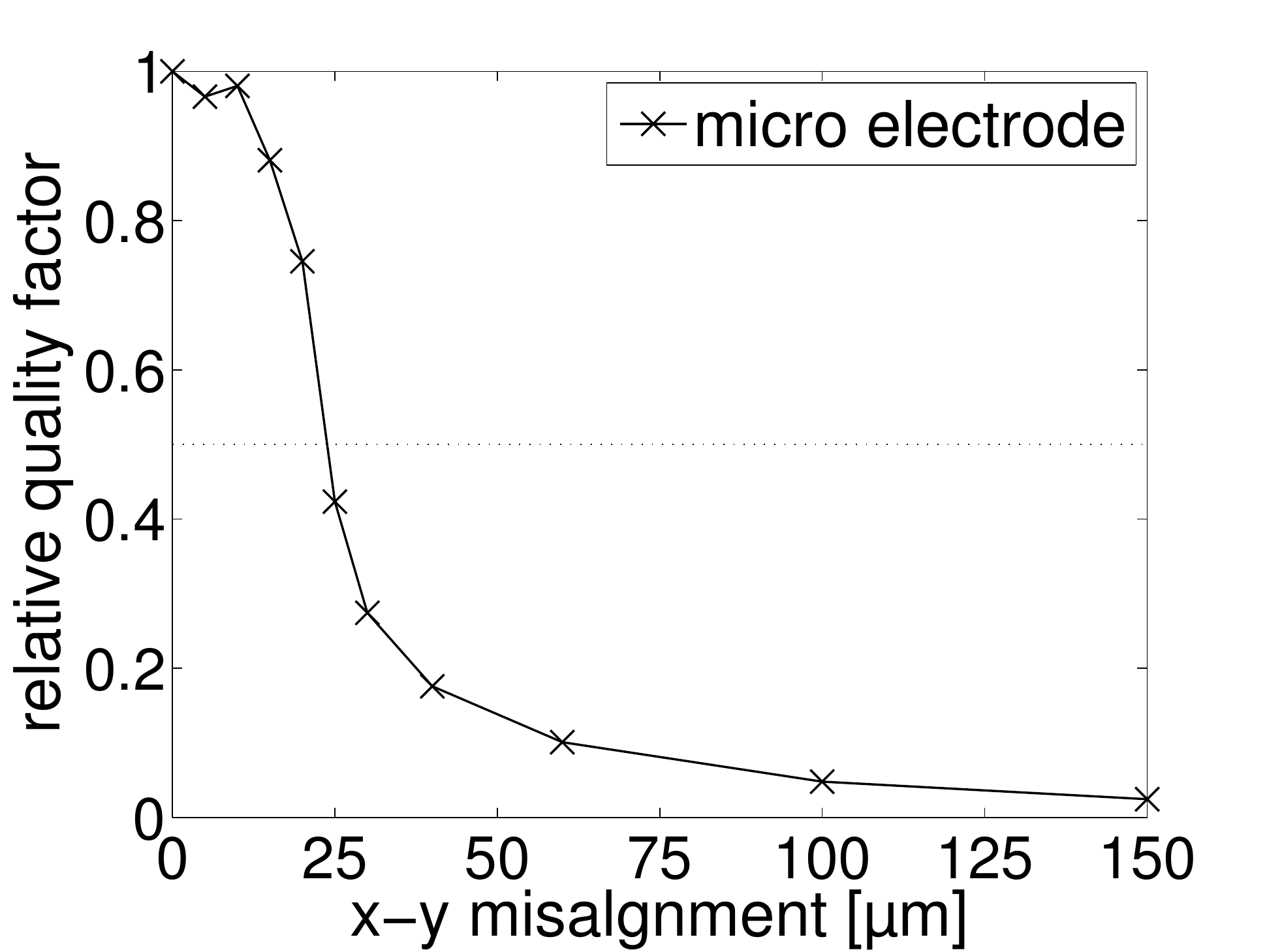}
	\end{minipage}
	\caption{\label{DataMis} Influence of alignment of the different degrees of freedom on the relative quality factor $\mathcal{F}/\mathcal{F_{\mbox{\scriptsize{max}}}}$. In each plot the dotted horizontal line represents 50 \% of the maximum of the quality factor. 
	A quantitative analysis is shown in table \ref{DataMisTable}.
	}
	\end{figure*}
\begin{figure}[ht]
	\includegraphics[width=0.5\textwidth]{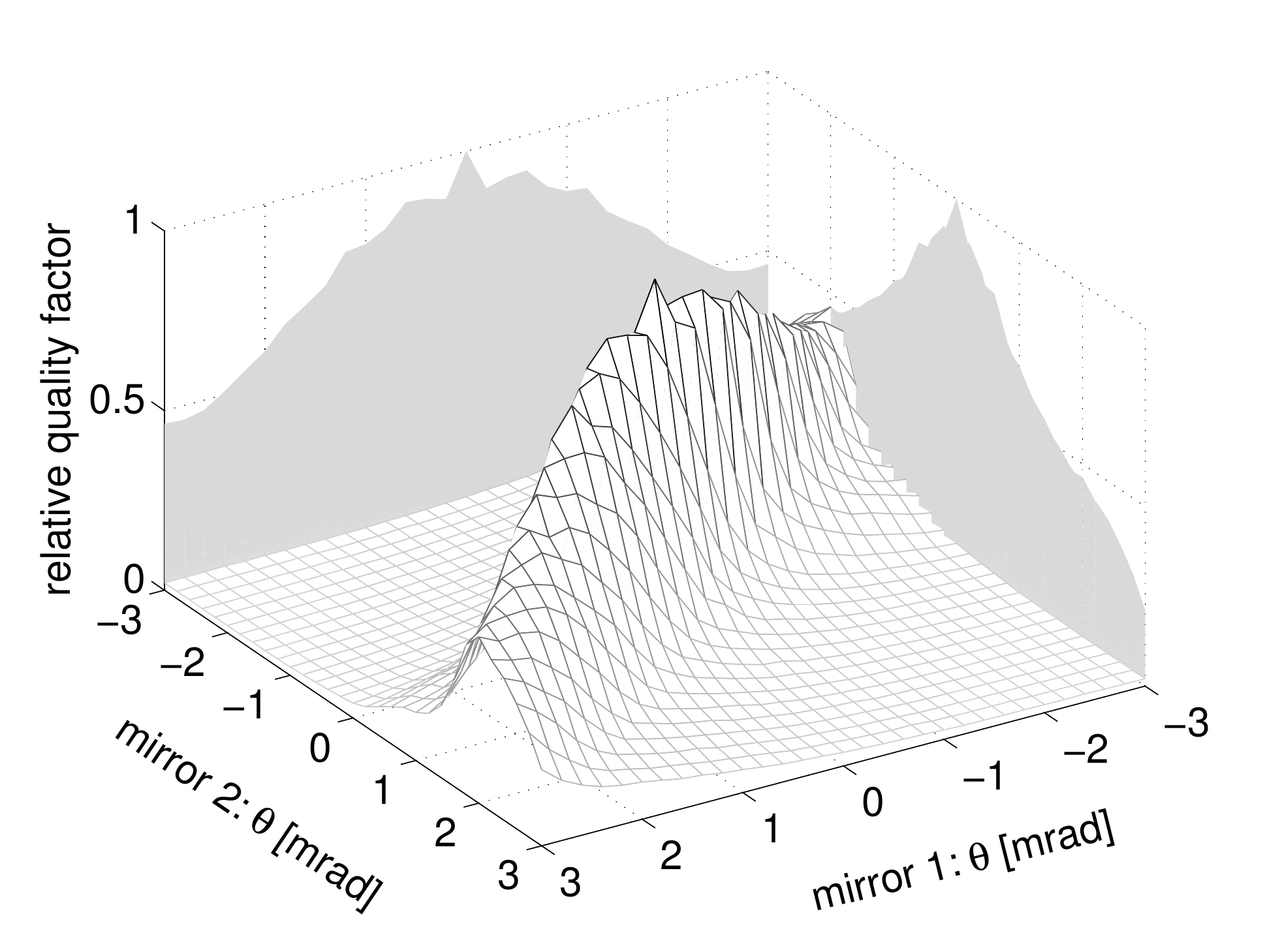}
	
	\caption{\label{SurfTheta} Relative quality factor plotted against the deviation from the optimum in the orientation of both mirrors.}
	\end{figure}

\begin{figure*}[ht]
	\begin{minipage}[ht]{0.3\textwidth}
	\flushleft
	\includegraphics[height=0.8\textwidth]{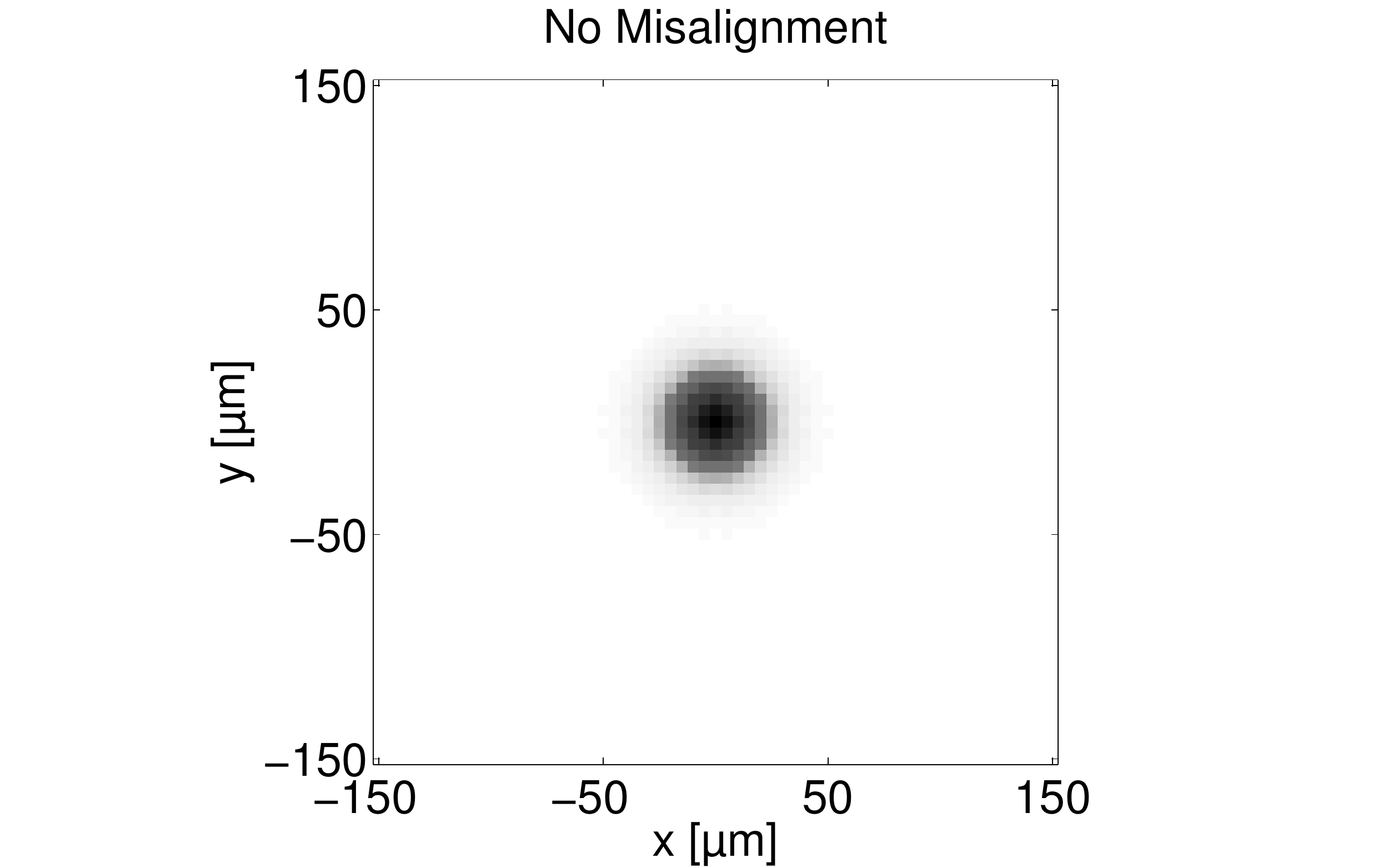}
	\end{minipage}
	\begin{minipage}[ht]{0.3\textwidth}
	\flushright
	\includegraphics[height=0.8\textwidth]{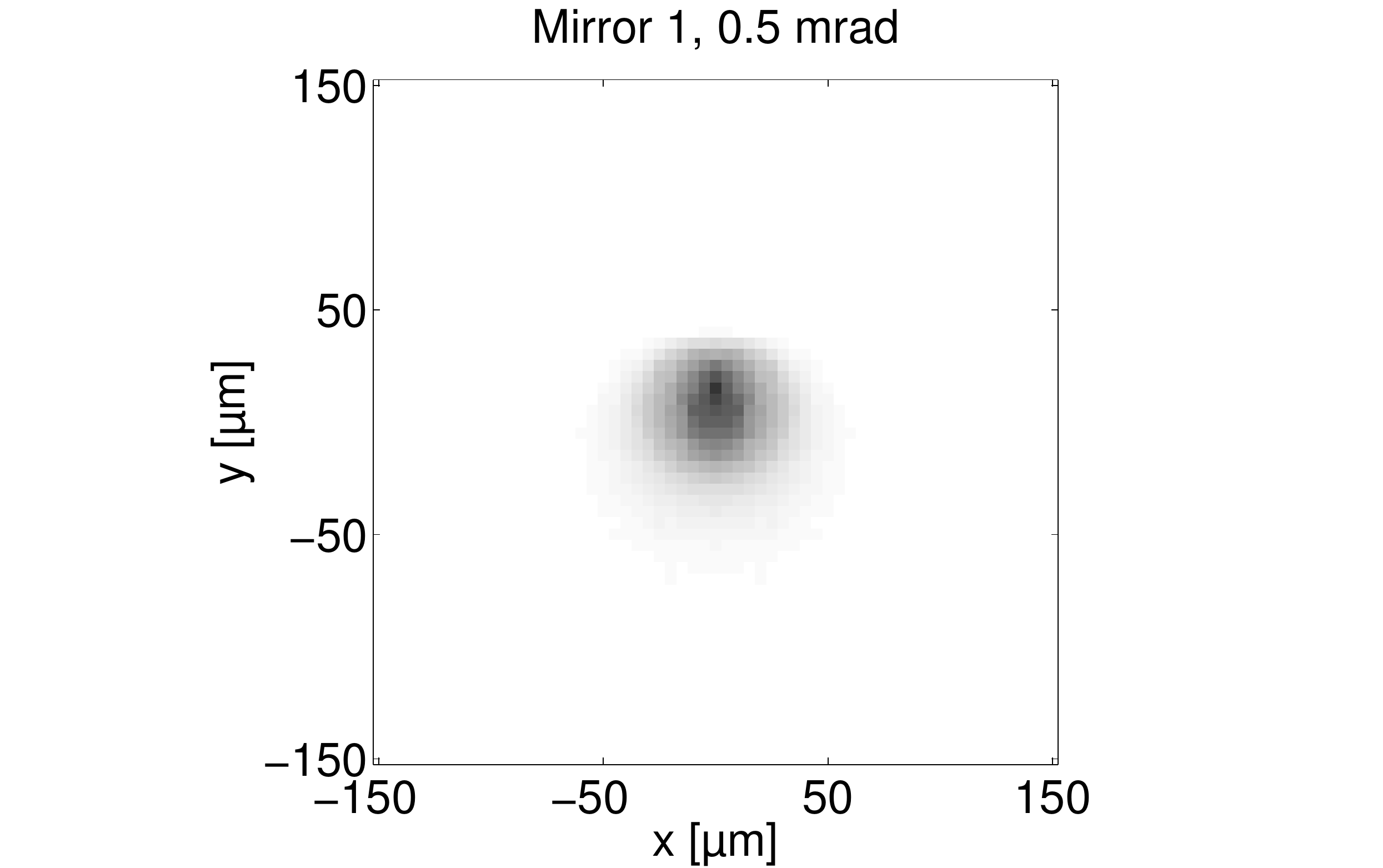}
	\end{minipage}
	\begin{minipage}[ht]{0.39\textwidth}
	\flushleft
	\includegraphics[height=0.6154\textwidth]{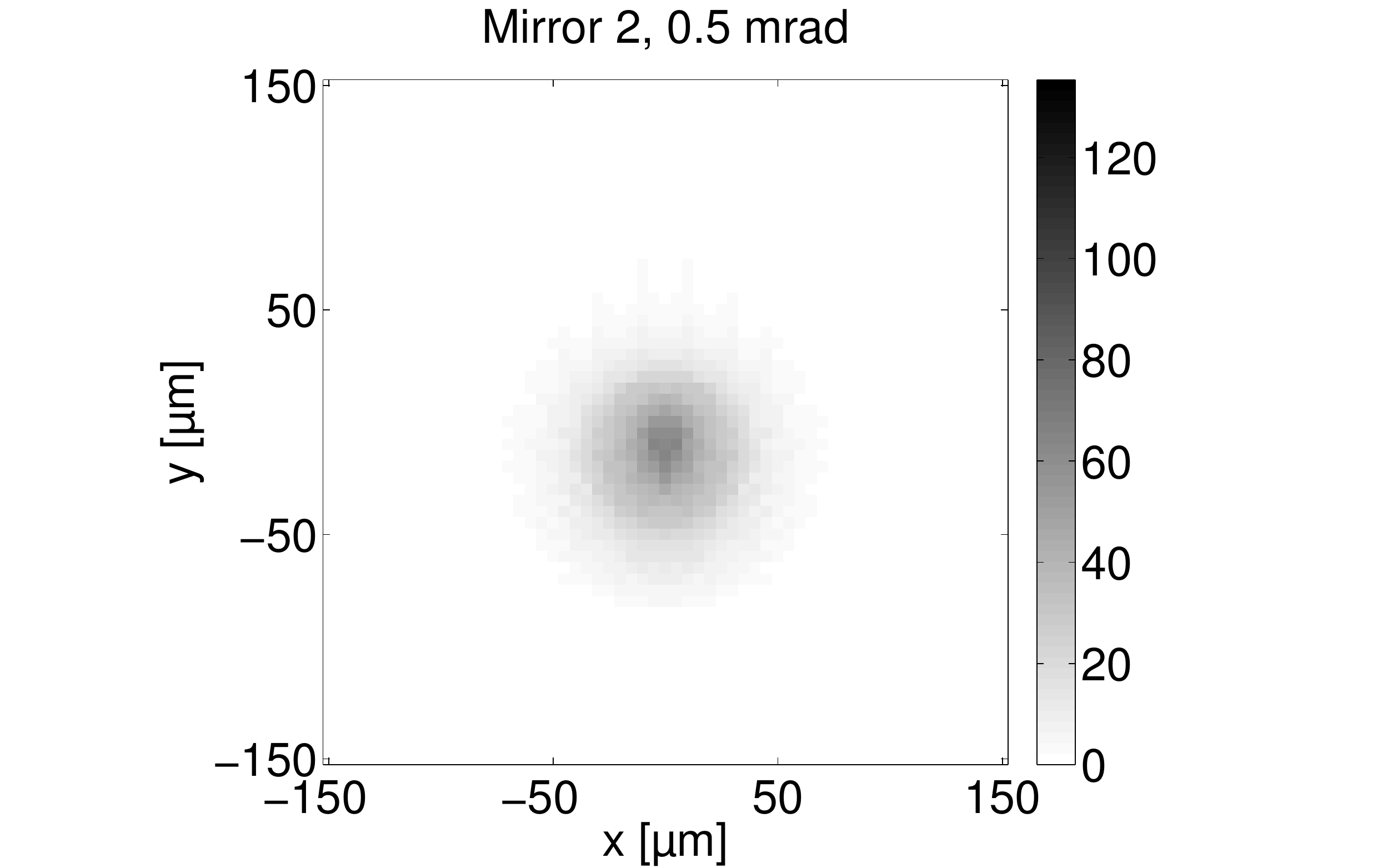}
	\end{minipage}\\
	\begin{minipage}[ht]{0.3\textwidth}
	\flushleft
	\includegraphics[height=0.8\textwidth]{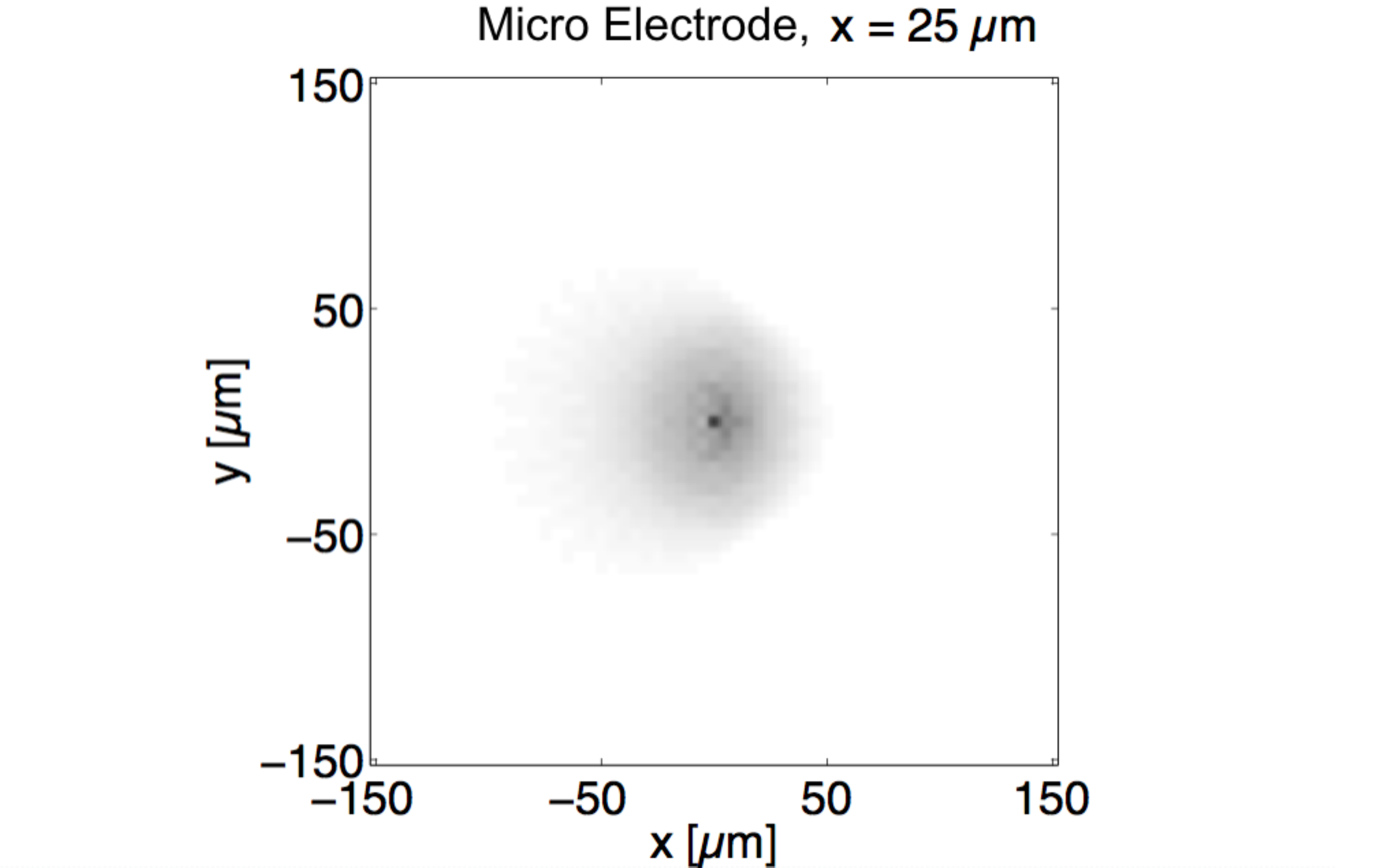}
	\end{minipage}
	\begin{minipage}[ht]{0.3\textwidth}
	\flushright
	\includegraphics[height=0.8\textwidth]{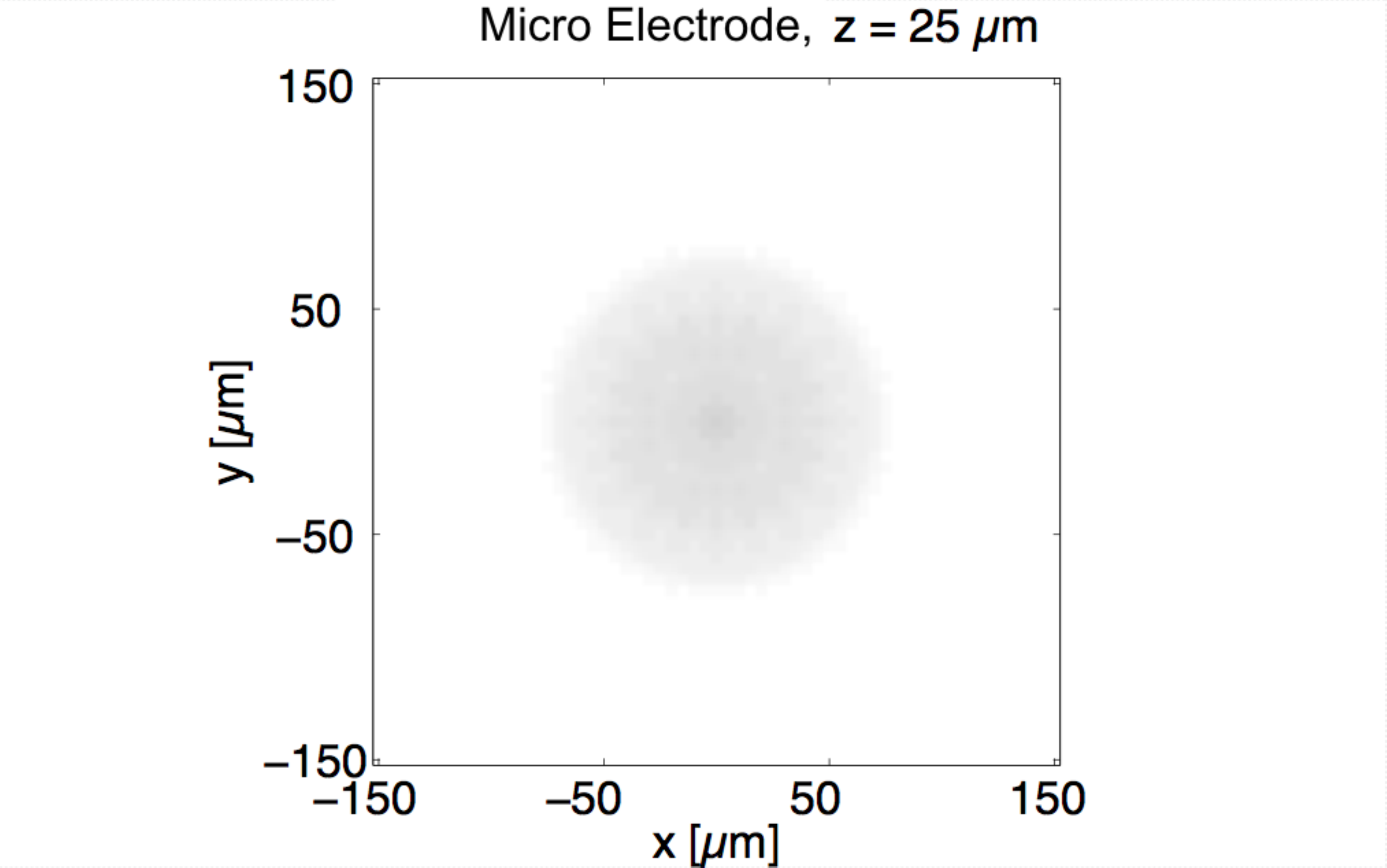}
	\end{minipage}
	\begin{minipage}[ht]{0.39\textwidth}
	\flushleft
	\includegraphics[height=0.6154\textwidth]{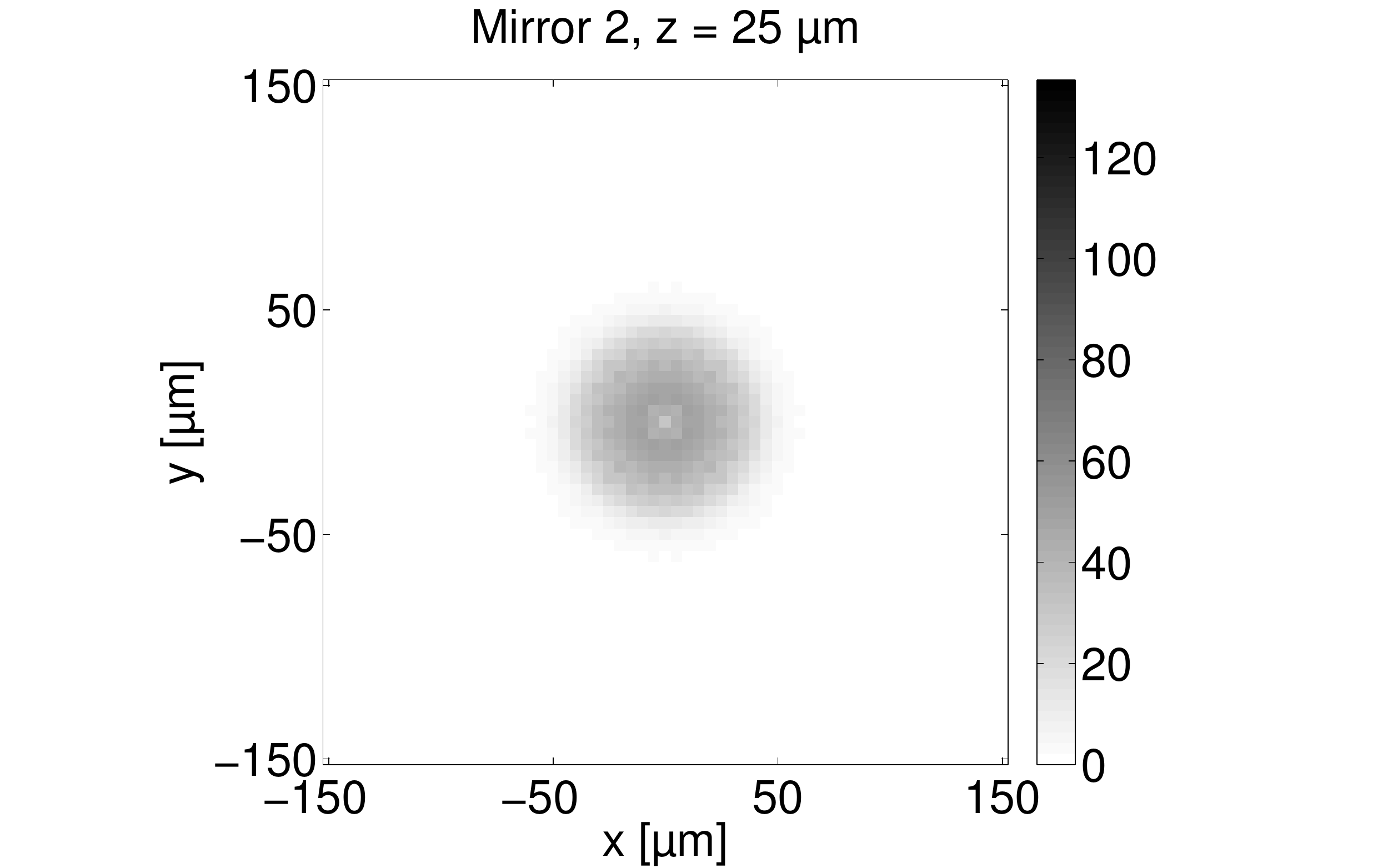}
	\end{minipage}
	\caption{\label{DataMisPic} Density distributions obtained with ray tracing simulations for different alignment scenarios: Upper row from left to right: no misalignment, mirror 1: $\Delta\theta=0.5$ mrad, mirror 2: $\Delta\theta=0.5$ mrad. Lower row from left to right: micro electrode: $\Delta x=25$ $\mu$m, micro electrode: $\Delta z=25$ $\mu$m, mirror 2: $\Delta z=25$ $\mu$m. Note, that the grey scale is the same in all plots (as indicated on the far right side in each row).}
	\end{figure*}

 \noindent
For the purpose of comparing the optical system to other possible optical setups (\textit{e.g.} based on lenses)  \cite{JINST} and quantifying the influence of possible misalignments, a ray-tracing code was developed. 
The homemade MatLab-based ray-tracing code takes into account the mirror geometries as well as the reflectivity of the mirror surfaces (using equations obtained in \cite{Coating}). 
To implement the micro electrode surface-dimensions, it is modeled as a collection of point-like light sources uniformly distributed over the 50 $\mu$m diameter active area of the micro electrode: The micro electrode surface is divided into 200 segments and for each segment a point-like light source is simulated, which is in the end weighted by the fraction the segment takes of the whole micro electrode surface. Each point-like light source is modeled as 1200 photons that are isotropically emitted in 4$\pi$. The pixel size of the detector amounts to (10$\times$10)$\mu$m$^2$ in the simulations.
\\
We introduce the quality factor $\mathcal F=\epsilon /\mbox{FAHM}$, to obtain quantitative results for the influence of misalignments, where FAHM denotes the full area at half maximum in mm$^2$ and $\epsilon$ the ratio of all the photons emitted from the light source to the photons contained in the FAHM. Because we are mostly facing asymmetric photon distributions when the optical elements are misaligned, we use the full area at half maximum instead of the full width at half maximum squared. 
The quality factor is proportional to the signal-to-background ratio $(S/N)=\mathcal F\cdot\big(R\cdot \epsilon_d \cdot 1/D\big)$, where R is the (photonic) isomeric decay rate, $\epsilon_d$ is the detector efficiency and D the dark count rate per area of the detector. 
The latter three values are all independent from misalignments, while $\mathcal F$ strongly depends on the alignment of each component (mirror 1, mirror 2 and the collection surface). The maximum value for $\mathcal F$ that is achieved with the ray tracing simulations of the optics is $\mathcal{F_{\mbox{\scriptsize{max}}}}=0.143 / (1.52\times10^{-3} \mbox{mm}^2)=94 /\mbox{mm}^2$.
The behavior of the quality factor as a function of misalignments of each degree of freedom is visualized in the plots in figs. \ref{DataMis} and \ref{SurfTheta}. 
The data of these plots was obtained by performing ray tracing simulations with the parameters described above and calculating $\mathcal{F}$ with the resulting photon density distributions. A collection of the photon density distribution plots is shown in fig. \ref{DataMisPic}.
To get a measure of the sensitivity of the alignment of each degree of freedom, the deviation at which the relative quality factor $\mathcal{F}/\mathcal{F_{\mbox{\scriptsize{max}}}}$ drops to 50\% or 80\%, respectively, is investigated by interpolating between neighboring data points. 
The results are shown in table \ref{DataMisTable}.  
	\begin{table*}[ht]
	\label{tab:1}   \centering    
	\begin{tabular}{c| c c c c c }
									 		&mirror 1: $\theta$ 	& mirror 2: $\theta$	&coll. electr.: z 	&mirror 2: z	&coll. electr.: x 	  \\
	\hline \\[-6pt]
	$\mathcal{F}/\mathcal{F_{\mbox{\scriptsize{max}}}}$ = 50\% 	 & 0.7 mrad			&0.4	 mrad			&9.7 $\mu$m			&20.8 $\mu$m		&23.8 $\mu$m \\ [2pt]
	$\mathcal{F}/\mathcal{F_{\mbox{\scriptsize{max}}}}$ = 80\%	 & 0.4 mrad			&0.2	 mrad			&4.5	$\mu$m		&13.5 $\mu$m		&18.0 $\mu$m \\ 
	\noalign{\smallskip}
	\end{tabular}
	\caption{Measures of the sensitivity of each degree of freedom on potential misalignments. The deviation at which the relative quality factor $\mathcal{F}/\mathcal{F_{\mbox{\scriptsize{max}}}}$ drops to 50\% or 80\%, respectively, is shown. The numbers were obtained by interpolating the data points shown in fig. \ref{DataMis}. \label{DataMisTable}}
	\end{table*}
The misalignment of two degrees of freedom at the same time may in some cases cancel out each other. The surface plot in fig. \ref{SurfTheta} shows the relative quality factor as a function of the orientation of both mirrors. There is clearly a ridge visible, where both misalignments cancel each other. In this way the drop in the signal-to-background ratio is damped, which can be seen in the shaded projections in fig. \ref{SurfTheta}. Depending on the alignment of the orientation of the other mirror, a significant drop in the signal-to-background ratio then appears only after a misalignment of more than 1 mrad. All in all, this gives a relatively larger parameter space for the alignment to end up with an acceptable signal-to-background ratio.\\ 
	 \noindent

	 \noindent
\subsection{Alignment}\label{AlignmentProc}

\begin{figure}[H]
	\includegraphics[width=0.5\textwidth]{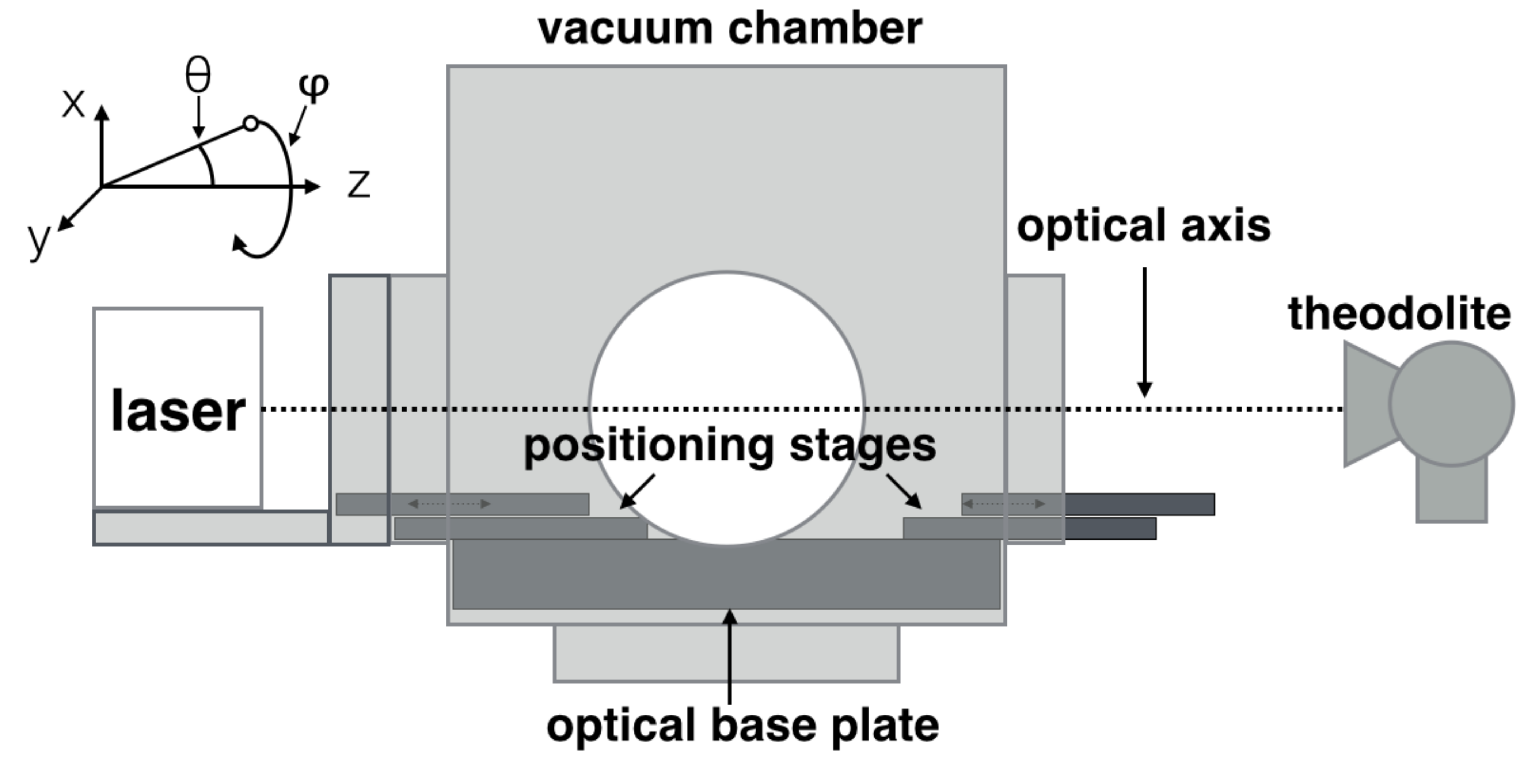}
	
	\caption{Alignment of the optical base plate and the positioning stages that are placed in the vacuum chamber. The adjustment laser and the theodolite are aligned with the symmetry axis of the chamber openings. \label{AlignmentIllustrationA} }
	\end{figure}
\begin{figure}[H]
	\begin{minipage}[ht]{0.24\textwidth}
	\flushleft
	\includegraphics[width=0.9\textwidth]{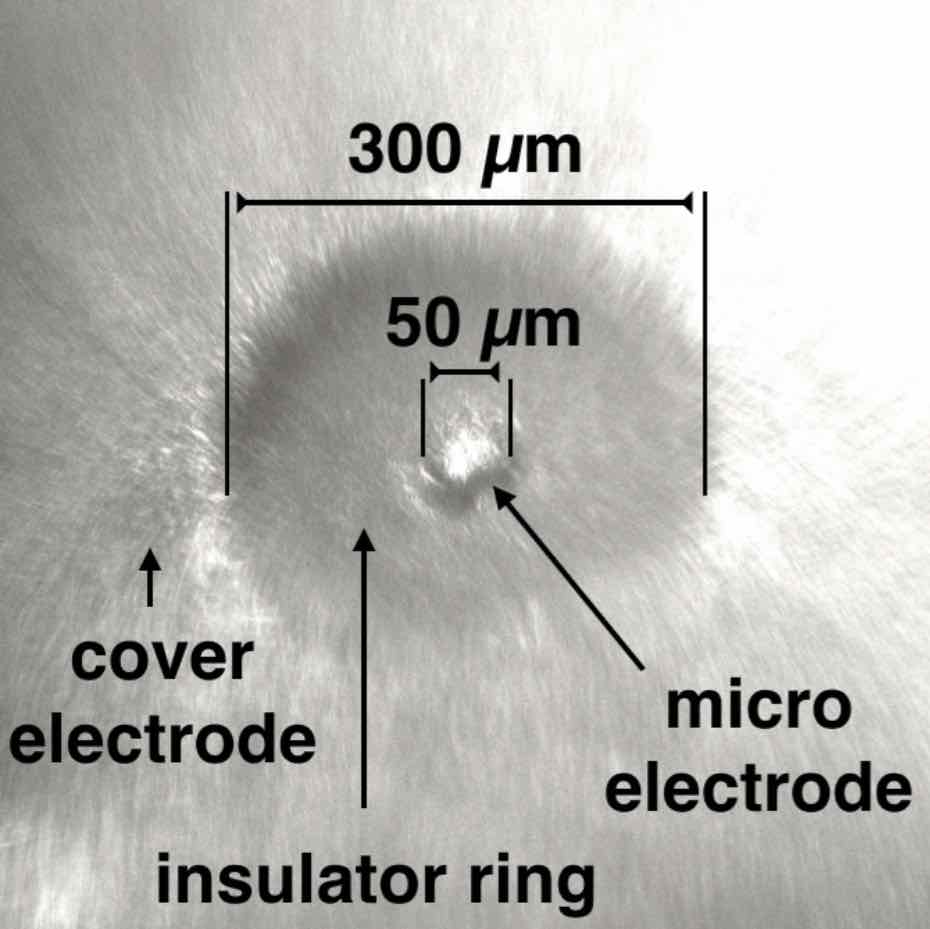}
	\end{minipage}
	\hfill
	\begin{minipage}[ht]{0.24\textwidth}
	\flushright
	\includegraphics[width=0.9\textwidth]{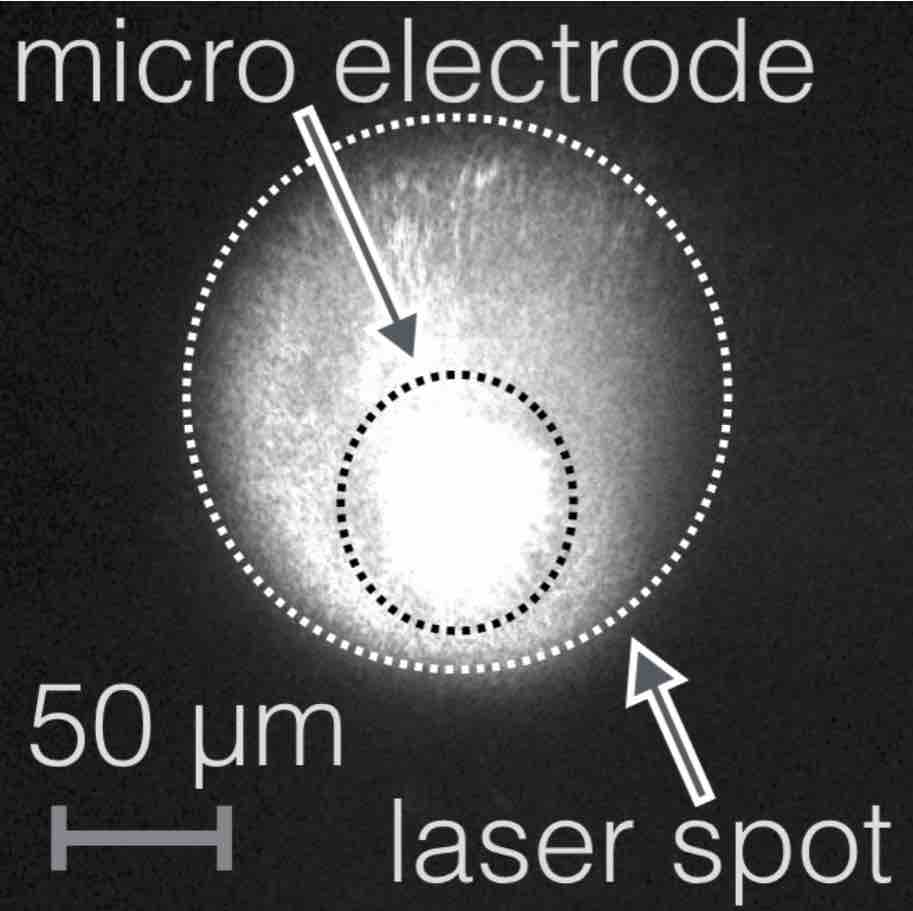}
	\end{minipage}
	\caption{\label{collectionMicroscope} Images of the collection surface taken with the microscopic optics. The left panel shows the micro electrode illuminated with a halogen lamp. The right panel displays an image of the laser spot focused onto the micro electrode.}
	\end{figure}

 \noindent
\textit{Optical axis and base plate:} First the optical axis is made visible with cross hairs that are spanned across the chamber opening flanges and an adjustment laser with an adjustable output power (\textit{Viasho, VA-I-N-532 (532 nm-1 300 mW)}) is aligned from the ion-beam side, while a theodolite (\textit{Zeiss, TH 2 M}) is aligned from the detector side (see fig. \ref{AlignmentIllustrationA}). Both mirrors are fixed by their mounting to the positioning stages. This enables to adjust their x-y positions by simply adjusting the x-y position of the stages that are mounted to the base plate. Therefore, two cross hair inserts are installed on the positioning stages. With their help, the optical base plate and thus the positioning stages with the mirrors are aligned relative to the previously aligned adjustment laser and theodolite.
	\begin{figure}[H]
	\includegraphics[width=0.5\textwidth]{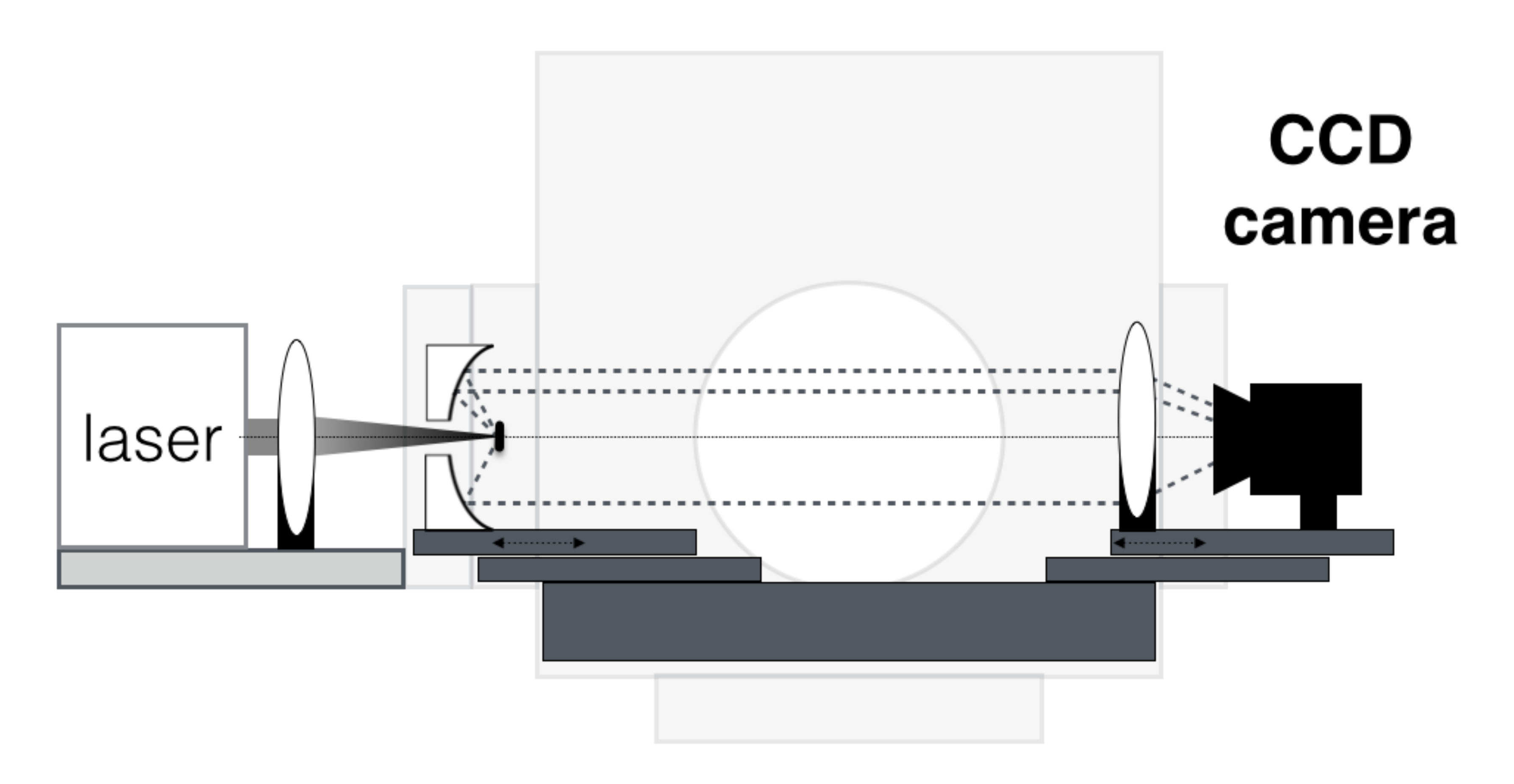}
	
	\caption{The collection surface is monitored with a microscopical optic that consists of a lens and a CCD camera (objective lens + CCD chip). \label{AlignmentIllustrationC} }
	\end{figure} \noindent
\textit{Z position of the micro electrode:} In order to roughly adjust the z position of the micro electrode with respect to mirror 1, the micro electrode is illuminated with a halogen lamp and observed with a microscopic optics, consisting of mirror 1, a lens (f=67 mm @ 532 nm) and a CCD camera. The focus of the arrangement consisting of only the lens and the CCD camera is set to infinity by focusing an object in about 10 m distance. 
In this way, the whole setup (mirror 1 + lens + CCD camera) only produces a sharp image when the micro electrode is in the (z-)focal spot of mirror 1. The setup (with a laser instead of the halogen lamp) is illustrated in fig. \ref{AlignmentIllustrationC}. Pictures that were taken with the microscope optics are shown in fig. \ref{collectionMicroscope}, where the micro electrode illuminated with a halogen lamp is shown in the left panel. 
	\begin{figure}[H]
	\includegraphics[width=0.5\textwidth]{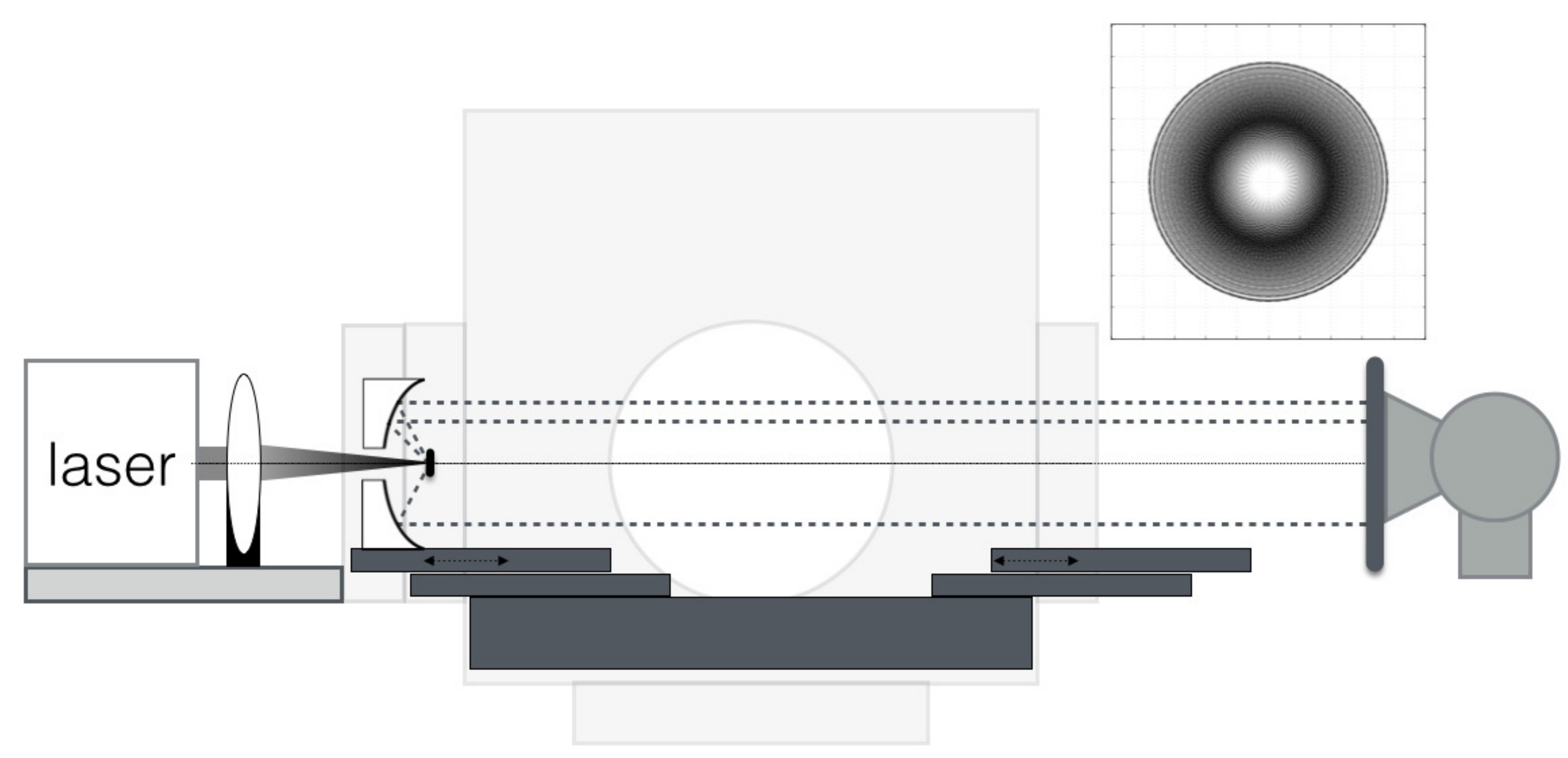}
	
	\caption{An image of mirror 1 is created on a screen that is attached to the theodolite, by focusing the laser with a lens onto the micro electrode. A simulation of the image that is created is shown in the upper right of the figure (see also fig. \ref{MisaligmentScenarios}). \label{AlignmentIllustrationB} }
	\end{figure} \noindent	
\\\textit{Alignment of mirror 1 and the micro electrode:}
In the next step, the previously aligned laser is focused on the collection surface with a lens (f = 67 mm @ 532 nm). The size of the laser spot on the micro electrode is monitored with the previously described microscope optics and minimized. The spot size is found to be smaller than 100 $\mu$m. By increasing the laser power and taking the microscope (lens + CCD camera) out of the beam path, it is possible to create an image on a screen mounted to the theodolite in about 2 m distance (see fig. \ref{AlignmentIllustrationB}). 
The image on the screen depends on several parameters that can be adjusted by in an iterative procedure: The goal is to obtain a shape of the image as a uniformly illuminated ring, centered around the optical axis. It is possible to fine tune the size of the ring by adjusting the z position of the collection surface. The shape of the image is influenced by the x-y position of the laser spot on the collection surface. The absolute position of the imaged ring on the screen is changed by adjusting the $\theta$ orientation of mirror 1.
Ray-tracing simulations of the image that is created on the screen, as well as the influence of a possible misalignment, are shown in fig. \ref{MisaligmentScenarios}.
Assuming the laser spot being aligned, the collection electrode is again monitored with the microscopic optics (see fig. \ref{AlignmentIllustrationC}) introduced above and aligned with the laser. A typical image of the laser spot as being aligned to the micro electrode is shown in the right panel of fig. \ref{collectionMicroscope}.
	 \noindent
\begin{figure}[H]
	\includegraphics[width=0.5\textwidth]{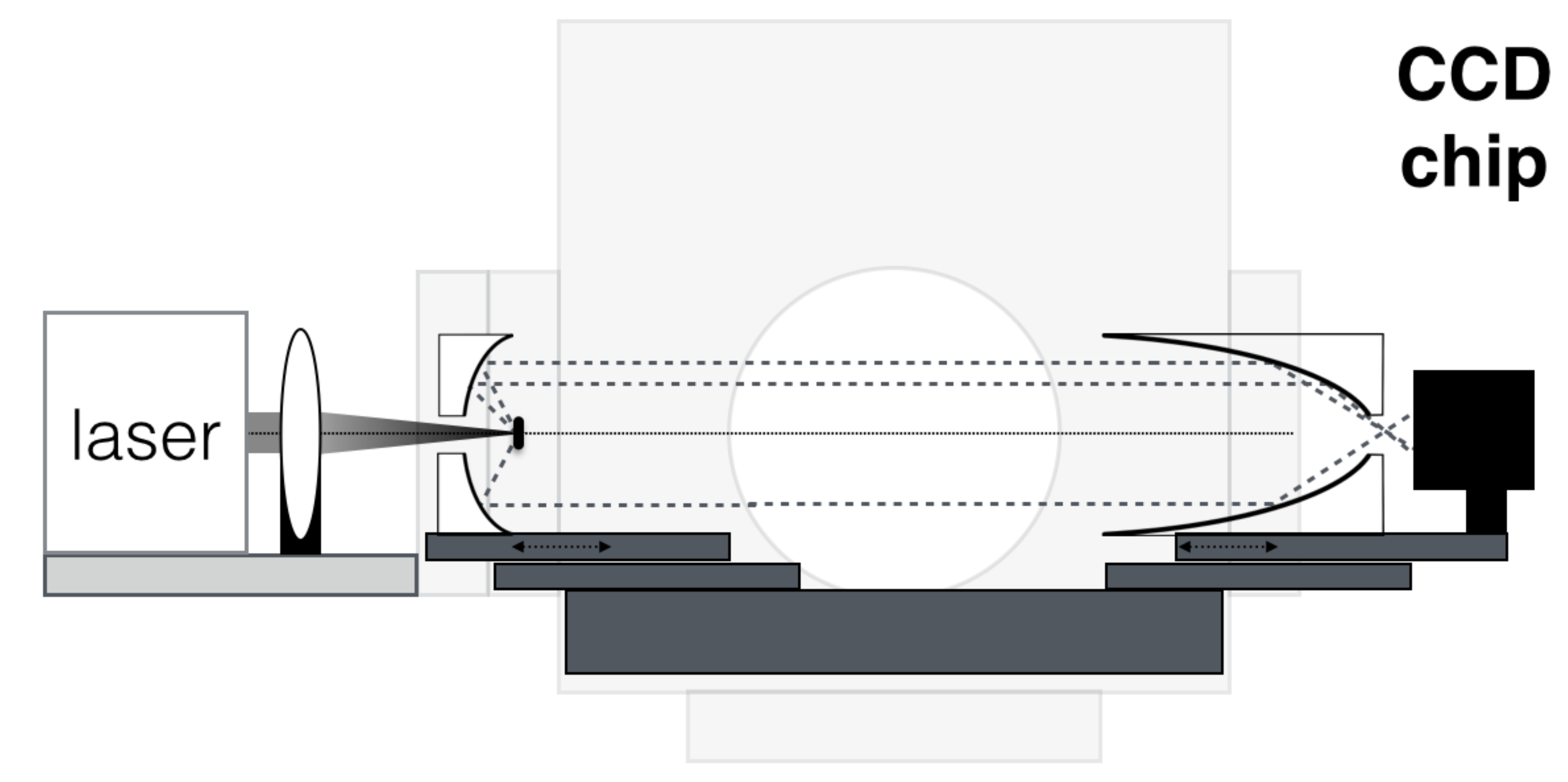}
	\caption{The light is imaged by mirror 2 onto a bare CCD chip. \label{AlignmentIllustrationD} }
	\end{figure} \noindent
\\\textit{Alignment of mirror 2:} The deep annular parabolic mirror is aligned by using the collimated light originating from the previously aligned mirror 1. To align the $\theta$ orientation of mirror 2, a bare CCD chip is placed about 3 mm behind the mirror exit (not in the focal spot of the mirror).
The setup is illustrated in fig. \ref{AlignmentIllustrationD}.  The image that is created is strongly sensitive on the $\theta$ orientation of mirror 2. 
For a perfectly aligned mirror, again a perfect ring is expected. 
Even a small tilt creates a heart-shaped structure on the detection plane, which can be used to align the $\theta$ orientation of mirror 2.
Since the CsI-coated MCP detector is only sensitive to wavelengths below $\approx$ 200 nm, the z alignment of mirror 2 needs to be performed with VUV light under vacuum conditions:
At first, the CCD chip is placed in the focal plane of mirror 2. 
Then the laser is replaced by a deuterium lamp, whose light is focused by two lenses onto the collection electrode. The deuterium lamp and the two lenses are aligned by recreating the image on the CCD chip that was previously obtained with the laser. Then the CCD chip is removed and the MCP detector is mounted onto the chamber opening. 
The adjustment of the z position of mirror 2 with respect to the MCP detector is now done by varying the distance between mirror 2 and the MCP detection plane and minimizing the spot size. 
It is then possible to save the position of the stepper motor with the LabView control program, remove the deuterium lamp and the lenses and finally connect the optics vacuum chamber to the extraction part of the setup.
%
%
\subsection{Precision achieved with the described alignment procedure}
\begin{figure*}[ht]
	\centering 
	\includegraphics[width=0.95 \textwidth]{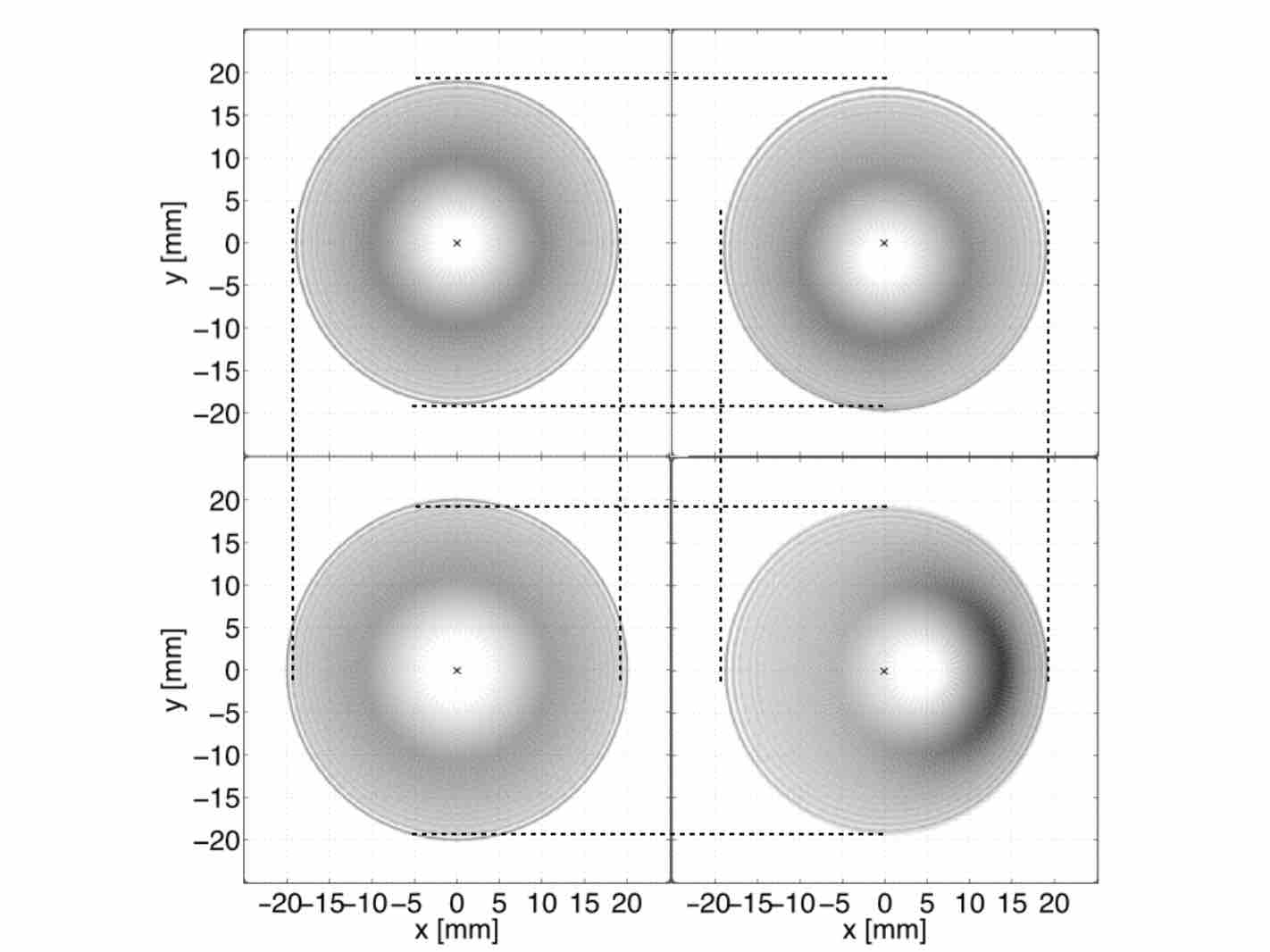}
	\caption{\label{MisaligmentScenarios} Simulation of misalignment scenarios that lead to the precisions given in table \ref{AlignTab}. The situation described in sect. \ref{AlignmentProc} ("\textit{Alignment of mirror 1 and the micro electrode}") is simulated. The density distributions on the screen mounted to the theodolite are shown. Upper left: without any misalignment, upper right: $\theta$ = 0.5 mrad misalignment of mirror 1, lower left: z = 10 $\mu$m of the micro electrode, lower right: x-y = 25 $\mu$m misalignment of the laser spot generated on the collection surface.}
\end{figure*}

In order to estimate the precision that is a\-chieved in the alignment procedure for each component, ray-tracing simulations of the alignment steps are performed and compared to the theoretically achievable optimum. 
For the alignment of mirror 1, where the screen is attached to the theodolite in 2 m distance to mirror 1, it is assumed that shifts of 1 mm from the optimum can be identified (\textit{i.e.} the image of mirror 1: 39 mm in diameter, perfect ring, central hole with 12 mm diameter).
A misalignment of $\theta$ leads to a shift $S= L \cdot \sin{\theta} \approx L \cdot \theta$, where $L$=2 m is the distance between mirror 1 and the screen on the theodolite. A precision of (1 mm)/(2 m) = 0.5 mrad for the $\theta$-alignment of mirror 1 is achieved. Simulations of different alignment scenarios are shown in fig. \ref{MisaligmentScenarios}. Each perturbation from the perfect image of mirror 1 is larger than 1 mm.
\begin{table}[h]\centering 
\begin{tabular}{p{.01 \textwidth} p{.15 \textwidth}  rr}
			&						&	\multicolumn{2}{ c}{\textbf{precision}}\\
			&						& \textbf{required} 		&	\textbf{achieved} \\ \hline
 \multicolumn{2}{ l}{\textbf{mirror 1}}			&			&			\\[4pt]
			&$\theta$-$\varphi$	[mrad]	& 0.7   	& 	0.5 \\[4pt]
		\hline
  \multicolumn{2}{ l}{\textbf{mirror 2}}			&	  		&	 \\ [4pt]
			&$\theta$-$\varphi$	[mrad]	& 0.4   		& 	0.5 \\
			&			z [$\mu$m]	& 21 		  	& 	30 \\ [4pt] \hline
 \multicolumn{2}{ l}{\textbf{micro electrode}}					&  		&	 \\ [4pt]
		&    			x-y [$\mu$m]		& 24		  	& 	15 \\
		&			z	[$\mu$m]		& 10 		 	& 	10 \\ [4pt] \hline
\end{tabular}
\caption{Summary of the required and achieved precisions for the alignment of the optical elements. \label{AlignTab}}
\end{table}
The same procedure has been performed in the alignment steps for mirror 2. The z alignment of mirror 2 is actually even limited by the resolution of the MCP detector in use. With the present  detector (25 $\mu$m channel diameter and assuming 50 $\mu$m pixel size), a precision of $\approx$ 30 $\mu$m is achieved.
The results of the precision analysis, as well as a comparison with the required precision obtained in sect. \ref{Ray-tracing} are shown in table \ref{AlignTab}.
The required precision for the alignment procedure is fulfilled for all degrees of freedom, except for the z position of mirror 2, which is even limited by the resolving power of the MCP detector.
Even if the required precision could be achieved in another way, in the end, the MCP detector could not resolve it, as it is the designated detector to measure the isomeric fluorescence radiation.

\section{Verification of the ion collection on the micro electrode}
\begin{figure}[H]
	\centering
	\includegraphics[width=0.5 \textwidth]{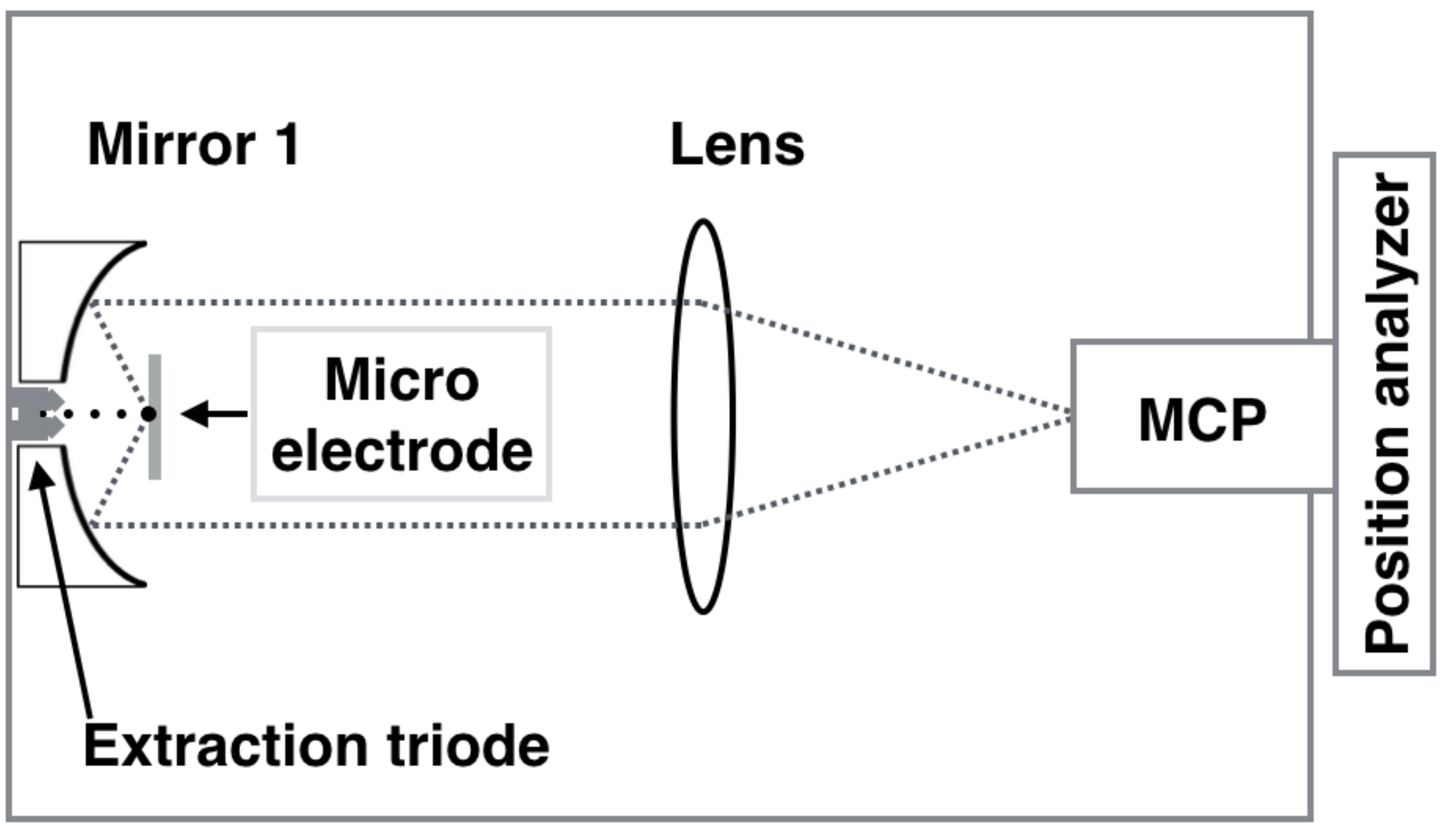}
	\caption{\label{setupCollection} Setup that was used to verify the ion collection on the micro electrode. The whole collection surface was phosphor coated (P11, ZnS:Ag) and $^{221}$Fr ions were accumulated on it. The light (emerging from $^{221}$Fr $\alpha$-decays) was parallelized by mirror 1 and focused onto the Quantar MCP surface by a CaF$_2$ lens. Different voltages were applied to the micro electrode and the cover electrode.}
\end{figure}
\begin{figure}[H]
	\centering
	\includegraphics[width=0.45 \textwidth]{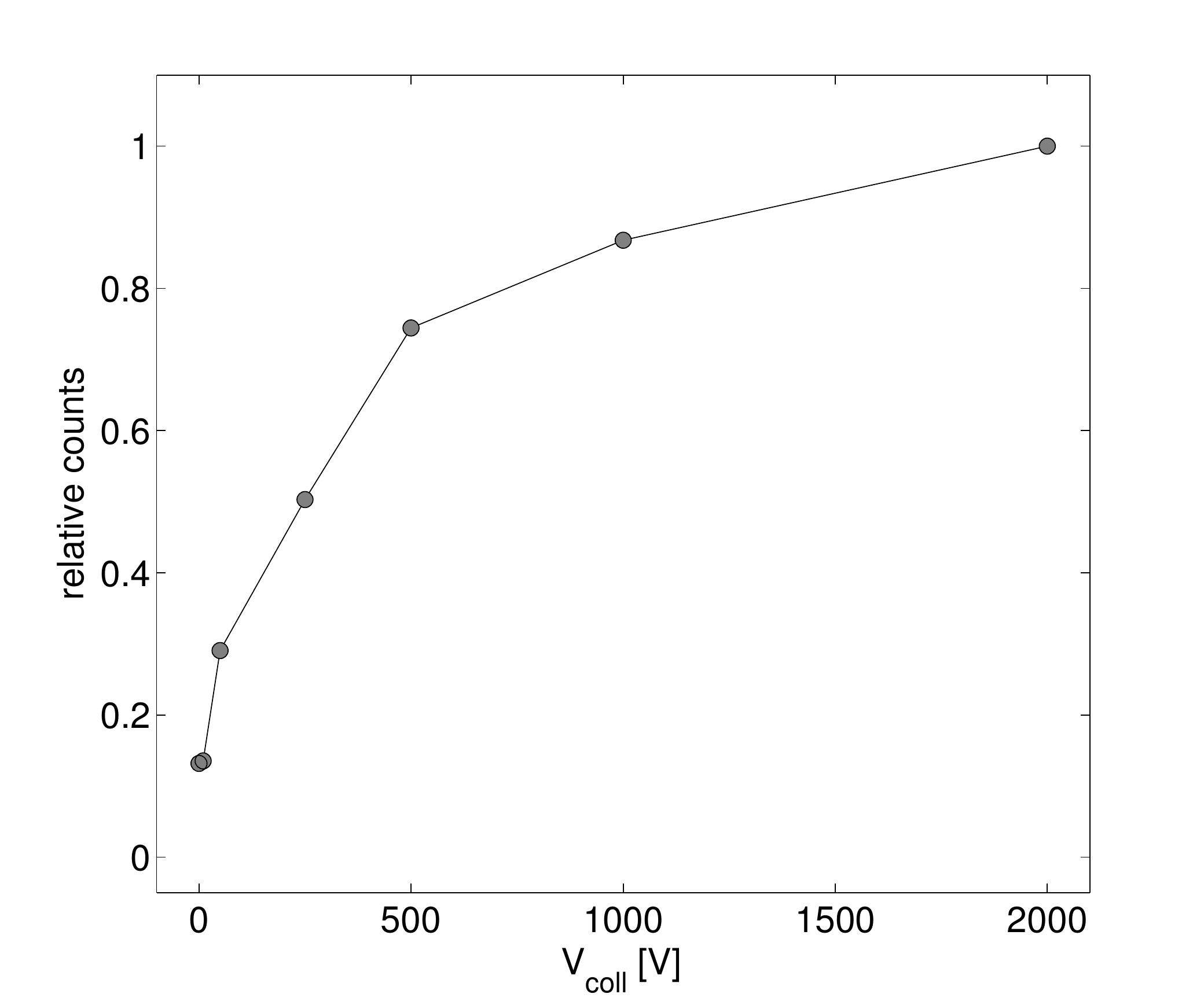}
	\caption{\label{DataCollectionPlot} Normalized count rate taken with the Quantar MCP plotted as a function of the voltage applied to the micro electrode. The data was obtained as described in the text and was normalized to the last data point. }
\end{figure}
\begin{figure*}[ht]
	\begin{minipage}[!ht]{0.33\textwidth}
	\begin{flushleft}\textbf{   ~~a}\end{flushleft}
	\includegraphics[height=0.69\textwidth]{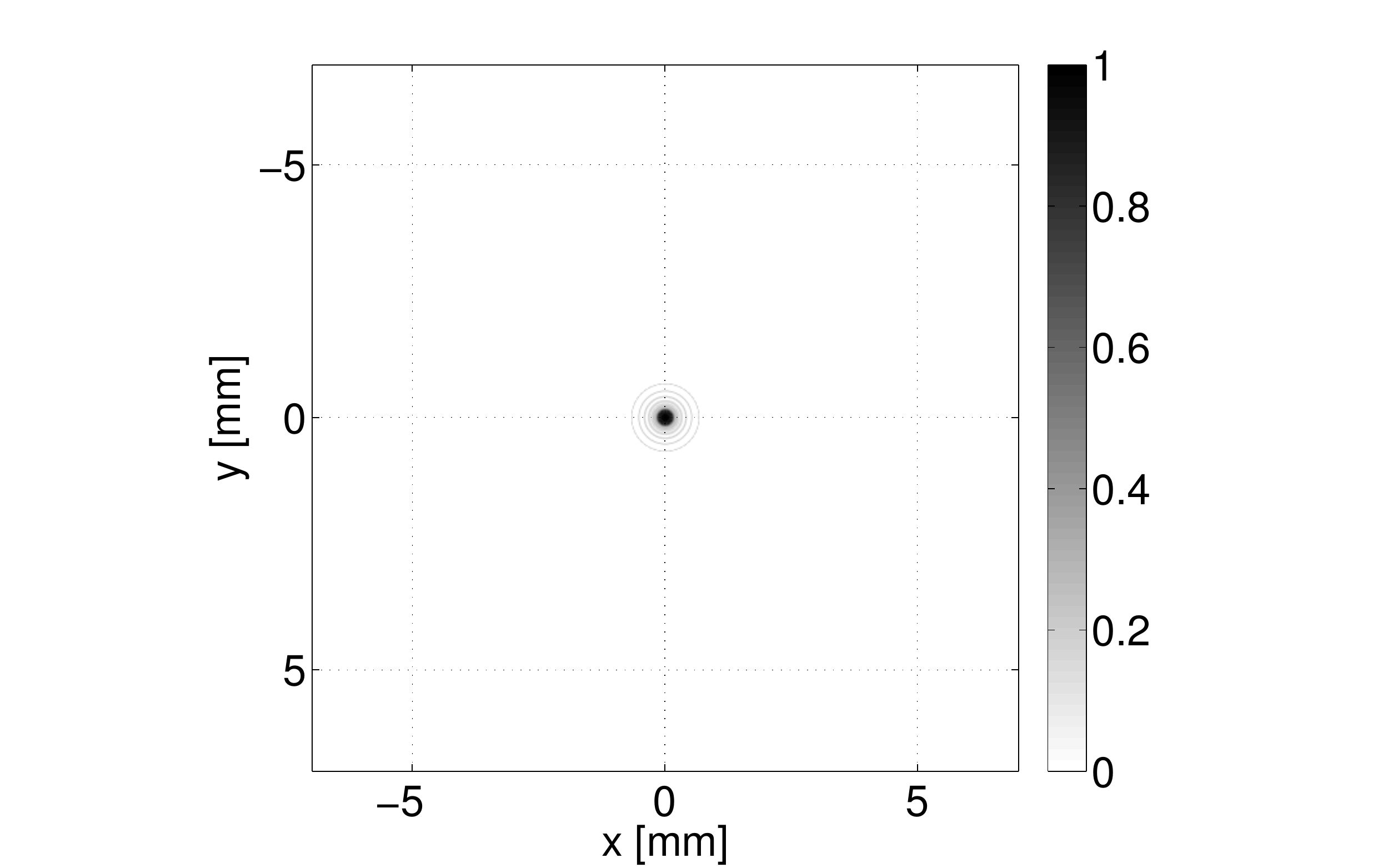} 
	\end{minipage}
	\begin{minipage}[!ht]{0.33\textwidth}
	\begin{flushleft}\textbf{~~~b}\end{flushleft}
	\includegraphics[height=0.7\textwidth]{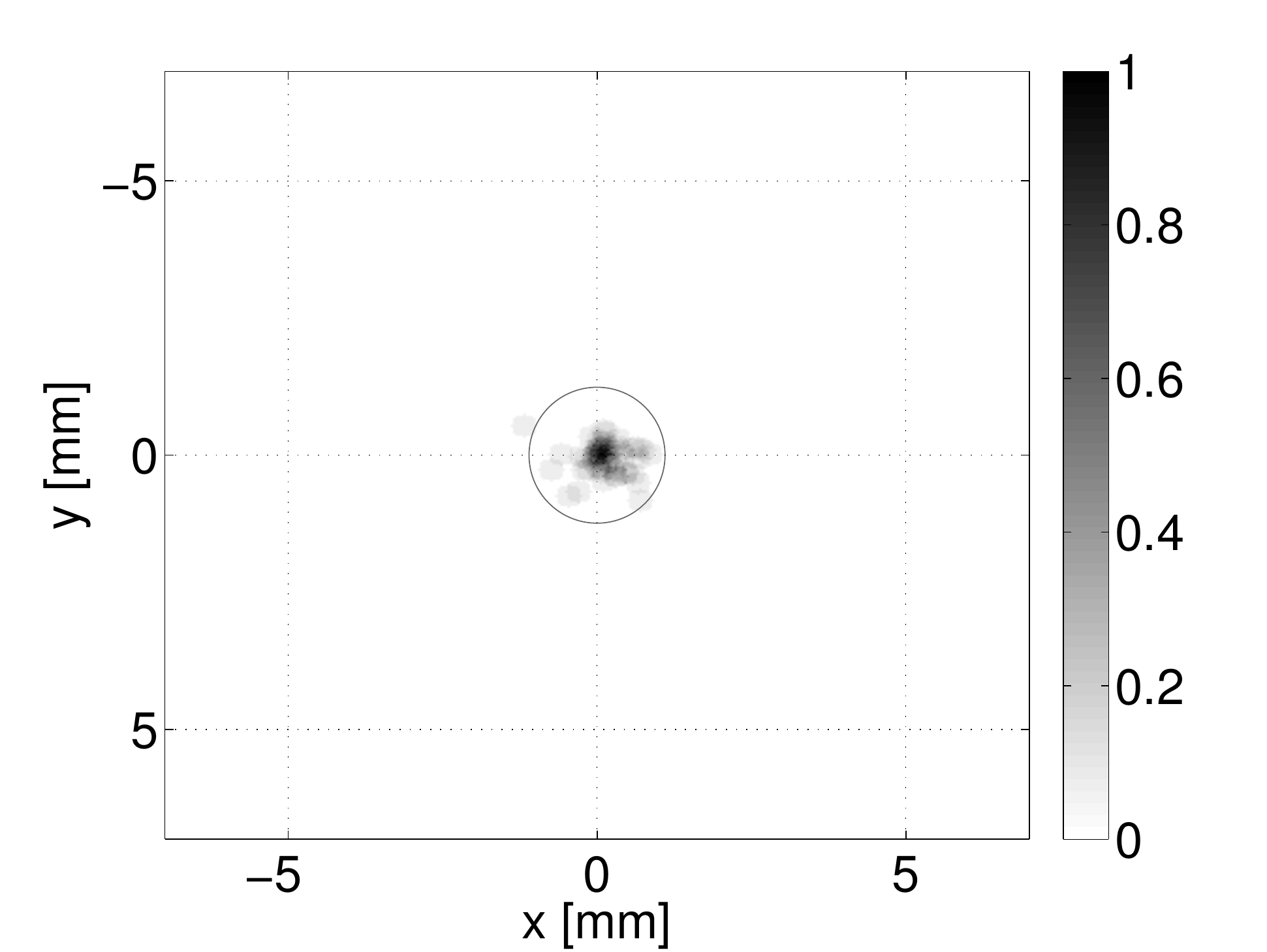} 
	\end{minipage}
	\begin{minipage}[!ht]{0.33\textwidth}
	\begin{flushleft}\textbf{~~c}\end{flushleft}
	\includegraphics[height=0.7\textwidth]{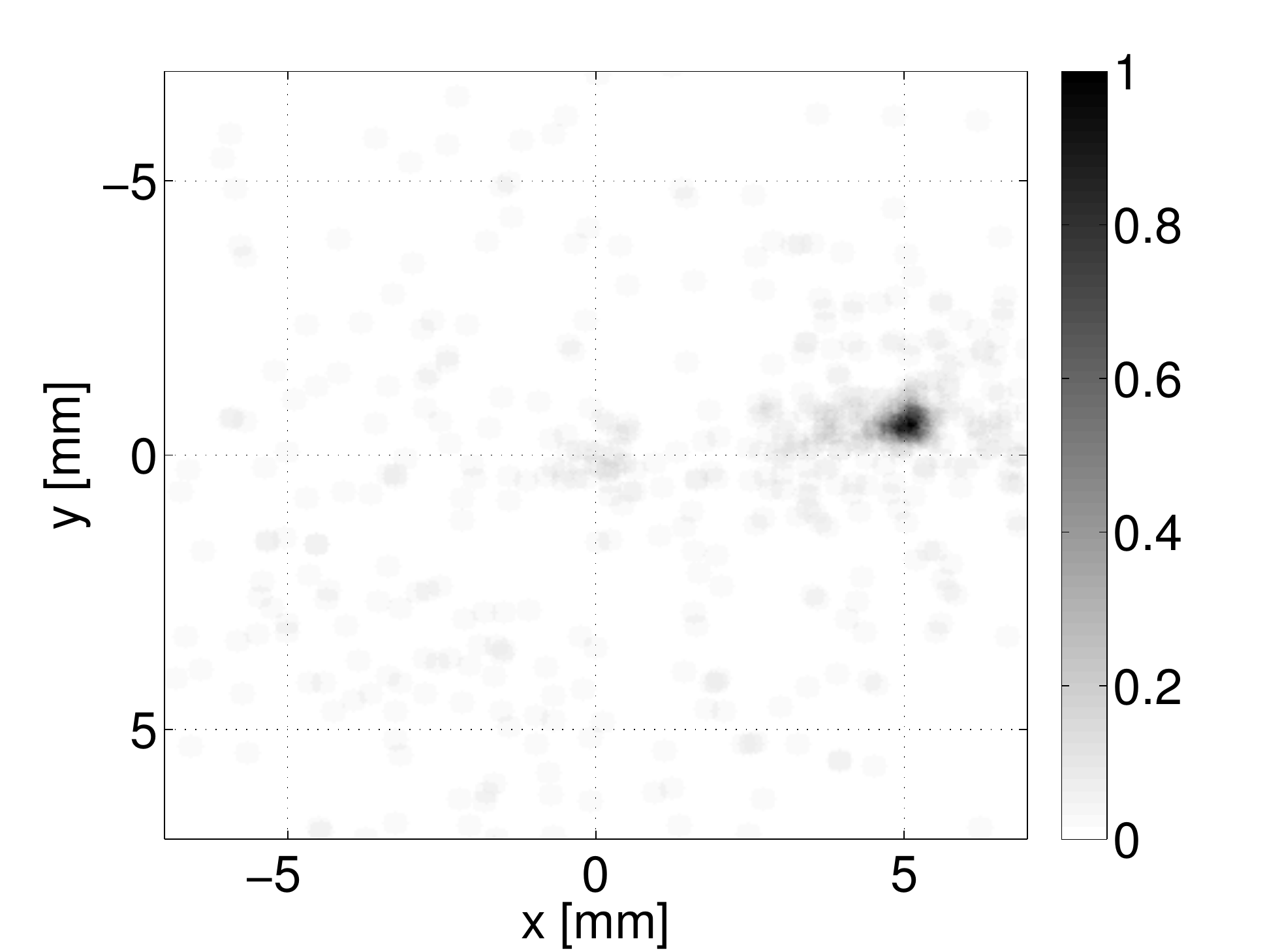} 
	\end{minipage}
	\caption{\label{CollFig} Plot of the photon distributions recorded with the setup shown in fig. \ref{setupCollection}. The left panel shows a simulation of the optics. The figure in the middle shows the measurement performed with $V_{\mbox{coll}}=-1000$ V and $V_{\mbox{cover}}=+3$ V at an MCP voltage of -2010 V. The circle marks the edge of the region where the entries were counted to obtain the data that is plotted in table \ref{DataCollectionPlot}. The right measurement was performed with $V_{\mbox{coll}}=V_{\mbox{cover}}=-1000$ V at an MCP voltage of -2500 V. Each plot is normalized to its maximum.}
\end{figure*}

The $^{229m}$Th isomer detection with the optical setup described above requires the collection of the thorium ions on the 50 $\mu$m micro electrode.
For the verification of the collection on the electrode, the whole collection surface was coated with a phosphor (P11 (ZnS:Ag) coating of 15 $\mu$m thickness, emitting light at a wavelength $\lambda$=460 nm) to localize radionuclides caught on the collection surface. Short-lived $\alpha$-decaying daughter nuclei from the $^{233}$U decay chain are accumulated on the electrode and the light emerging from the $\alpha$-decays on the collection surface is detected. The presented measurements were performed while collecting $^{221}$Fr$^{2+}$ ($t_{1/2}$=4.9 min \cite{NNDC}).
The setup is shown in fig. \ref{setupCollection} and consists of mirror 1, a lens  of a focal length of f=106 mm at 160 nm wavelength and a single-photon imaging system MCP (\textit{Quantar, Model 2601B Mepsicron-II Single-Photon Imaging Detector System} \cite{Quantar}) with a built-in position analyzer. The quantum efficiency of the system is above 12\% at 400 nm.
For a single measurement, 150 frames, each with an exposure time of 4 s, were recorded. 
To obtain count rate values, the number of entries in a circle of a radius of 1.2 mm around the peak-center position was counted. The position and dimensions of the circle are visualized in fig. \ref{CollFig}. 
A measurement series was performed by varying the voltage applied to the micro electrode $V_{\mbox{coll}}$ (from 0 V to -2000 V), leaving the cover electrode at $V_{\mbox{cover}}=$+3 V. 
The resulting data, normalized to the maximum count rate, is visualized in fig. \ref{DataCollectionPlot} which clearly shows the dependency of the obtained signal on the applied potential.
The corresponding 2D photon distributions are shown in fig. \ref{CollFig}.
The measurement series results in a saturation curve, which complies with the expected behavior of the ions responding to the different voltages applied to the micro electrode, see fig. \ref{DataCollectionPlot}.
The middle panel shows a measurement performed with $V_{\mbox{coll}} = -1000$ V, while $V_{\mbox{cover}}=+3$ V. The distribution (with a width of $\approx$0.6 mm) shown in fig. \ref{CollFig}\textbf{b}, corresponds to the $\alpha$-decay of $^{221}$Fr on the \O\xspace=50 $\mu$m micro electrode. Fig. \ref{CollFig}\textbf{a} shows the corresponding simulation of the optical setup exhibiting the ideal photon distribution.
Further data was taken with the cover- and the micro electrode at the same potential (-1000 V), shown on the right panel of fig. \ref{CollFig}\textbf{c}. 
Unlike as in fig. \ref{CollFig}\textbf{b}, the spot in fig. \ref{CollFig}\textbf{c} is at a different position, since, when both electrodes are at the same potential, the ions are accelerated in the z direction onto the surface and do not get deflected by any field gradients in the x-y direction. 
In this way it could be verified, that the ions react on the voltages applied to the micro electrode. 
Therefore, for an efficient ion collection, the positioning of the micro electrode with respect to the extraction-triode exit does not require a high precision and it is sufficient to place the micro electrode in the center of mirror 1 in order to gain the maximum light yield.
\section{Verification of the optical setup properties in the VUV spectral range}
\begin{figure*}[ht]
	\centering
	\includegraphics[width=0.95 \textwidth]{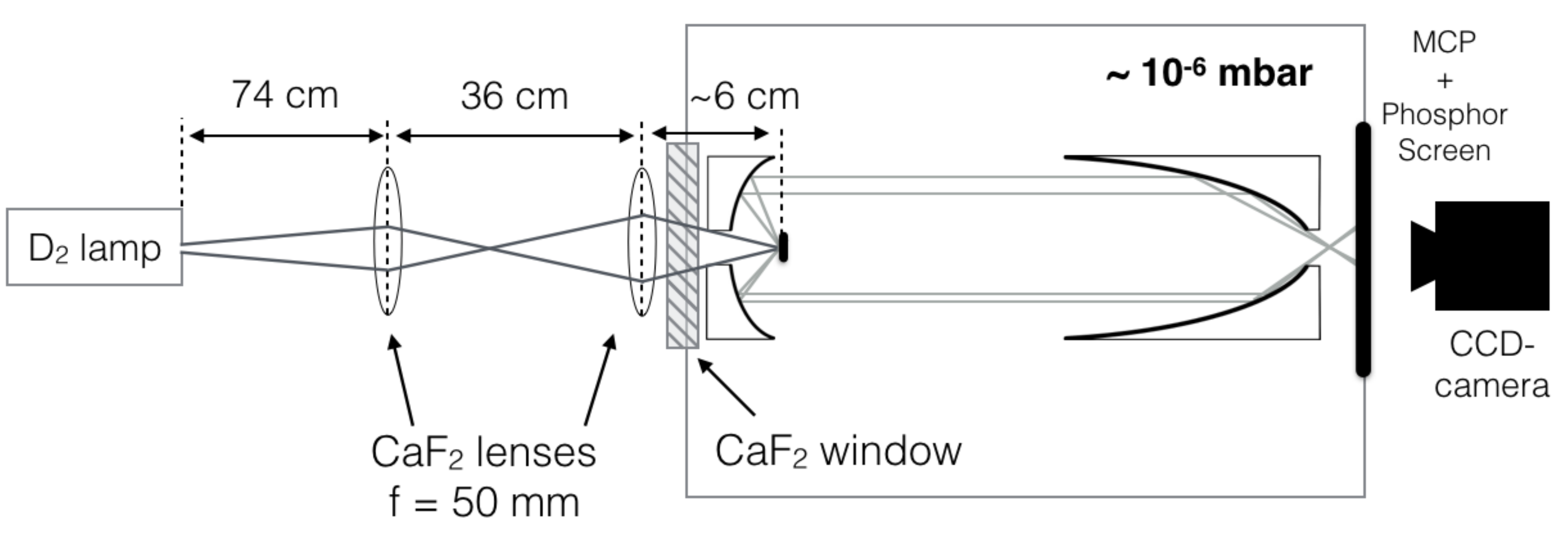}
	\caption{\label{TestSetup} Sketch of the setup that was used to test the optical properties in the VUV spectral range. The light emitted from a D$_2$ lamp is focused with two CaF$_2$ lenses onto the collection electrode, where it is scattered. The scattered light mimics a point-like light source on the micro electrode.}
\end{figure*}
\begin{figure}[H]	

	\begin{minipage}[ht]{0.45\textwidth}
	\flushleft
	\includegraphics[width=1\textwidth]{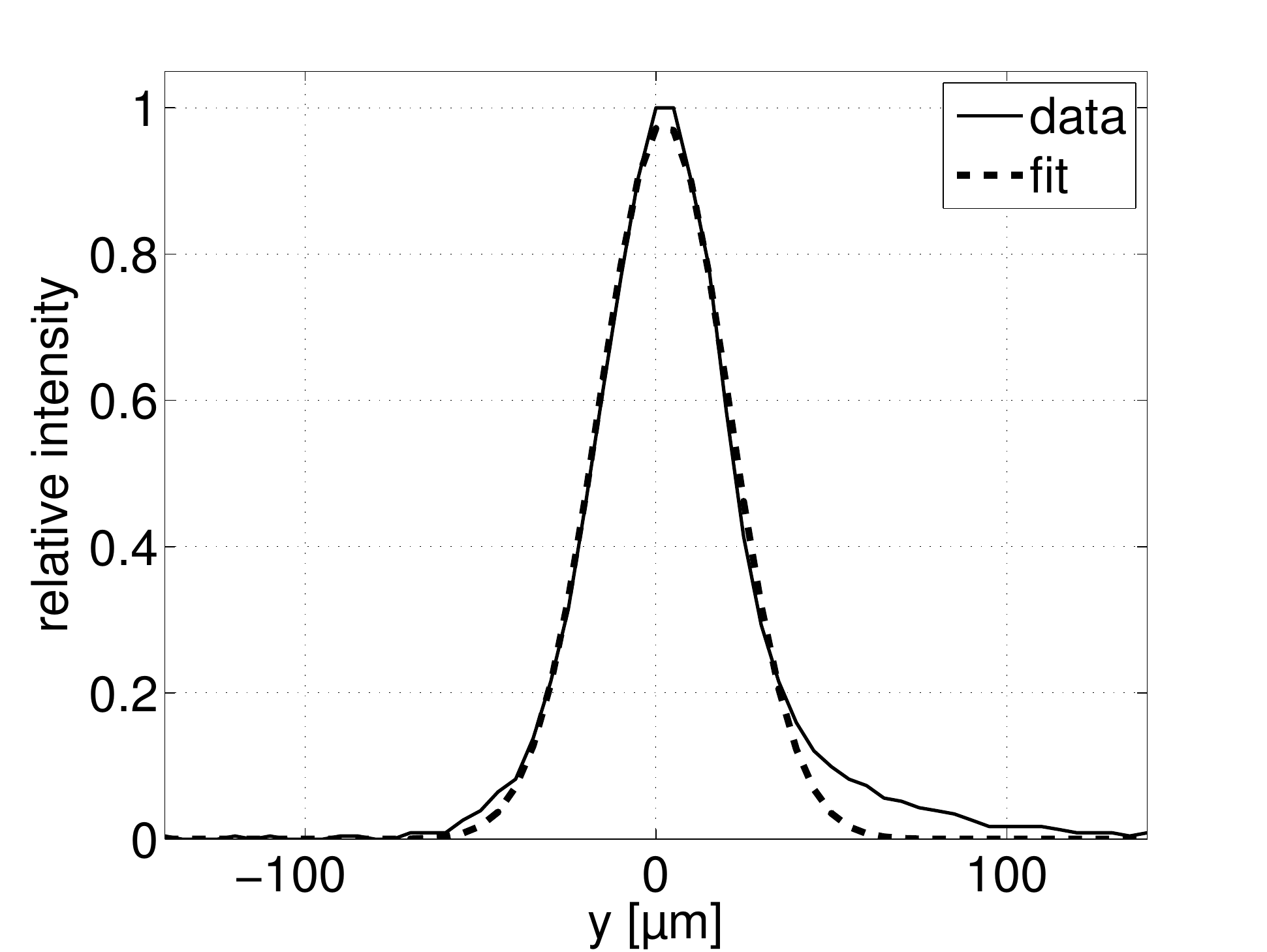}
	\end{minipage}
	\caption{\label{collectionVUV} D$_2$ lamp output aperture ($\O$=1mm) imaged onto the CCD chip with two lenses (f=50 mm at 160 nm), to probe the achievable spot size on the micro electrode. The profile of the image is shown. The FWHM was measured to be 44 $\mu$m, the full width at 10\% of maximum (FWTM) was measured to be 80 $\mu$m.}
\end{figure}

The idea behind the optical test measurements presented here is to mimic a light source located on the micro electrode by generating a small radiating spot on the electrode.
For the test measurements, a deuterium (D$_2$) lamp (\textit{Heraeus VUV light source, V05 30 W}) was used. It features two emission maxima at 120 nm and 160 nm, respectively. Since it does not emit collimated light, two lenses were used to generate a relatively small focal spot on the micro electrode. The setup is illustrated in fig. \ref{TestSetup}. The VUV lamp and the two CaF$_2$ lenses (f=50 mm at 160 nm) are placed outside the vacuum chamber. A CaF$_2$ window marks the entrance to the vacuum chamber, where the optical setup, described in sect. \ref{VUV optics}, is placed. The MCP phosphor screen is monitored by a CCD camera (\textit{Pointgrey Flea 2 FL2- 14S3M-C}, with built-in CCD chip \textit{Sony ICX267 CCD}) with an objective lens (\textit{25mm Computar M2514 MP2 Megapixel Lens C-Mount}) and a 10 mm spacer ring. The whole field of view is 10 $\times$ 7.38 mm$^2$.
It should be mentioned that the light emitted from the VUV lamp propagates through more than 1 m of air, therefore the spectral shape of the light detected by the MCP is not well defined. 
Since the properties of the all-reflective optical setup remain nearly constant above a wavelength of about 130 nm, the exact wavelength of the detected light is not of special importance, as long as being detectable by the MCP detector.
A measurement of the spot size was performed by imaging the D$_2$-lamp output aperture via two lenses onto the bare CCD chip (at atmospheric pressure).
The CCD chip was placed in a distance of about 6 cm to the last lens, which corresponds to its distance to the micro electrode in the complete test setup, see fig. \ref{TestSetup}.
Fig. \ref{collectionVUV} shows the spot size that is achieved with the two lenses on the micro electrode. The FWHM lies below 50 $\mu$m (the full width at 10\% of maximum (FWTM) is 80 $\mu$m) and thus is qualified to simulate point like radiation on the micro electrode. In the paraxial approximation, the magnification created by the two lenses is $\approx$ 70:1. This factor is not reached in the measurements, due to chromatic and spherical aberrations of the lenses. Instead a maximum magnification of $\approx$ 23:1 is achieved.
 \noindent
\begin{figure}[]
	\centering	
	\includegraphics[width=0.45\textwidth]{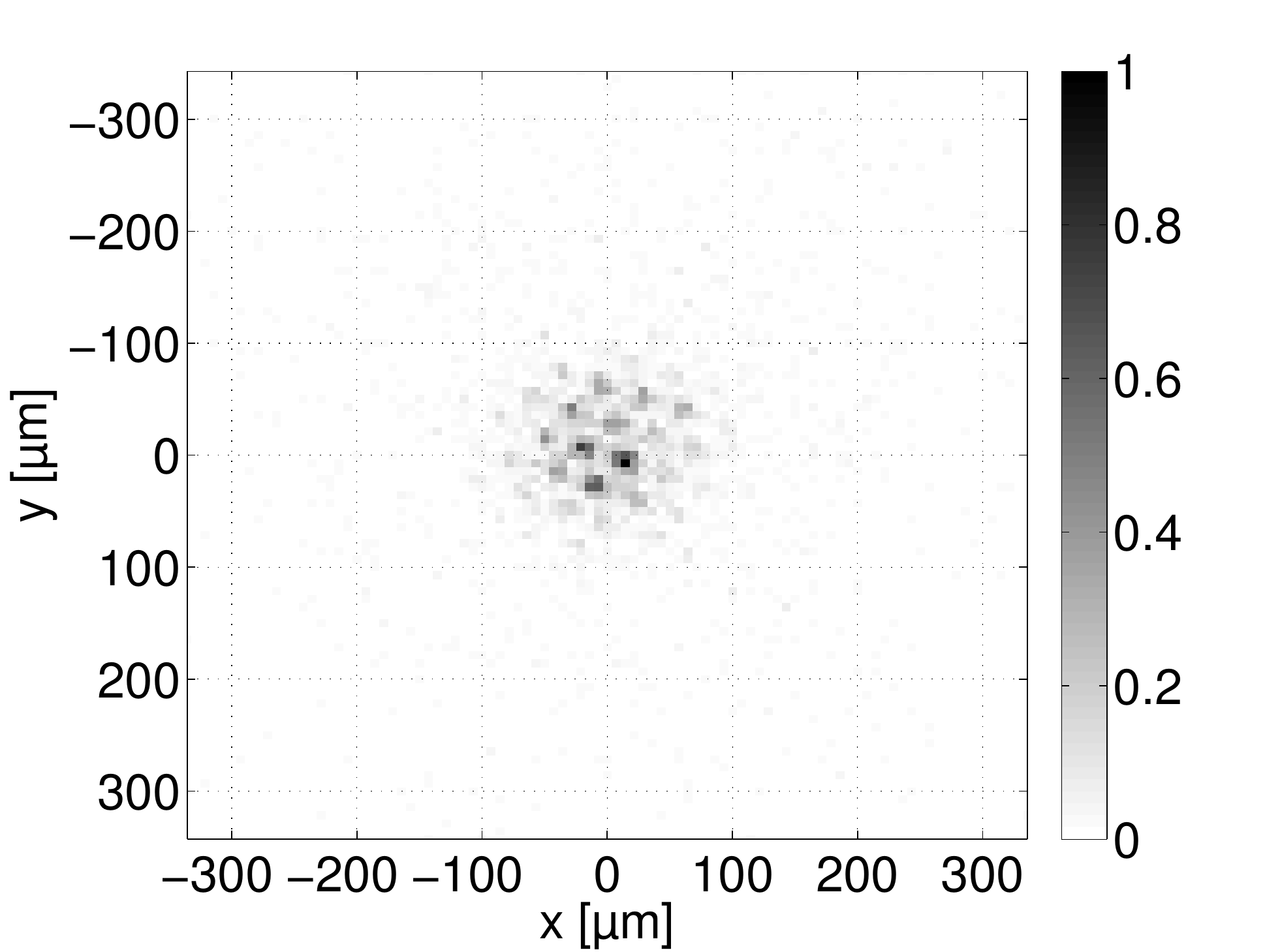}
	\caption{\label{Discrete} Plot of the peak centroid positions resulting from the analysis described in the text. A discrete distribution is clearly visible, which results directly from the resolved MCP channels.}
\end{figure} \noindent
\begin{figure*}[ht]
\begin{minipage}[hb]{0.45\textwidth}
\begin{flushleft}\textbf{~~a}\end{flushleft}
	\centering
	\includegraphics[height=0.8\textwidth]{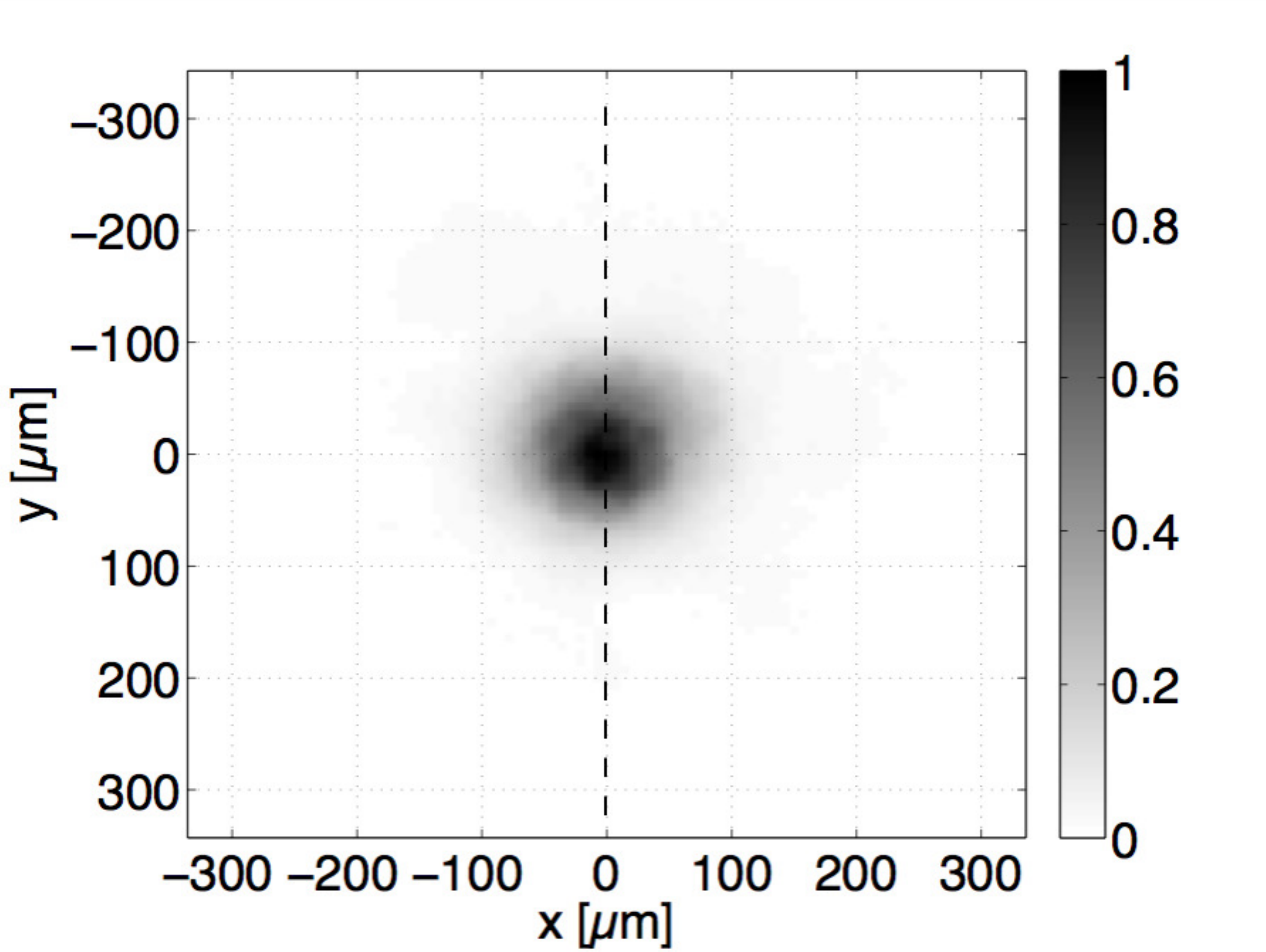}
	\end{minipage}
	\begin{minipage}[ht]{0.45\textwidth}
	\begin{flushleft}\textbf{~~~~~~~~b}\end{flushleft}
	\centering
	\includegraphics[height=0.8\textwidth]{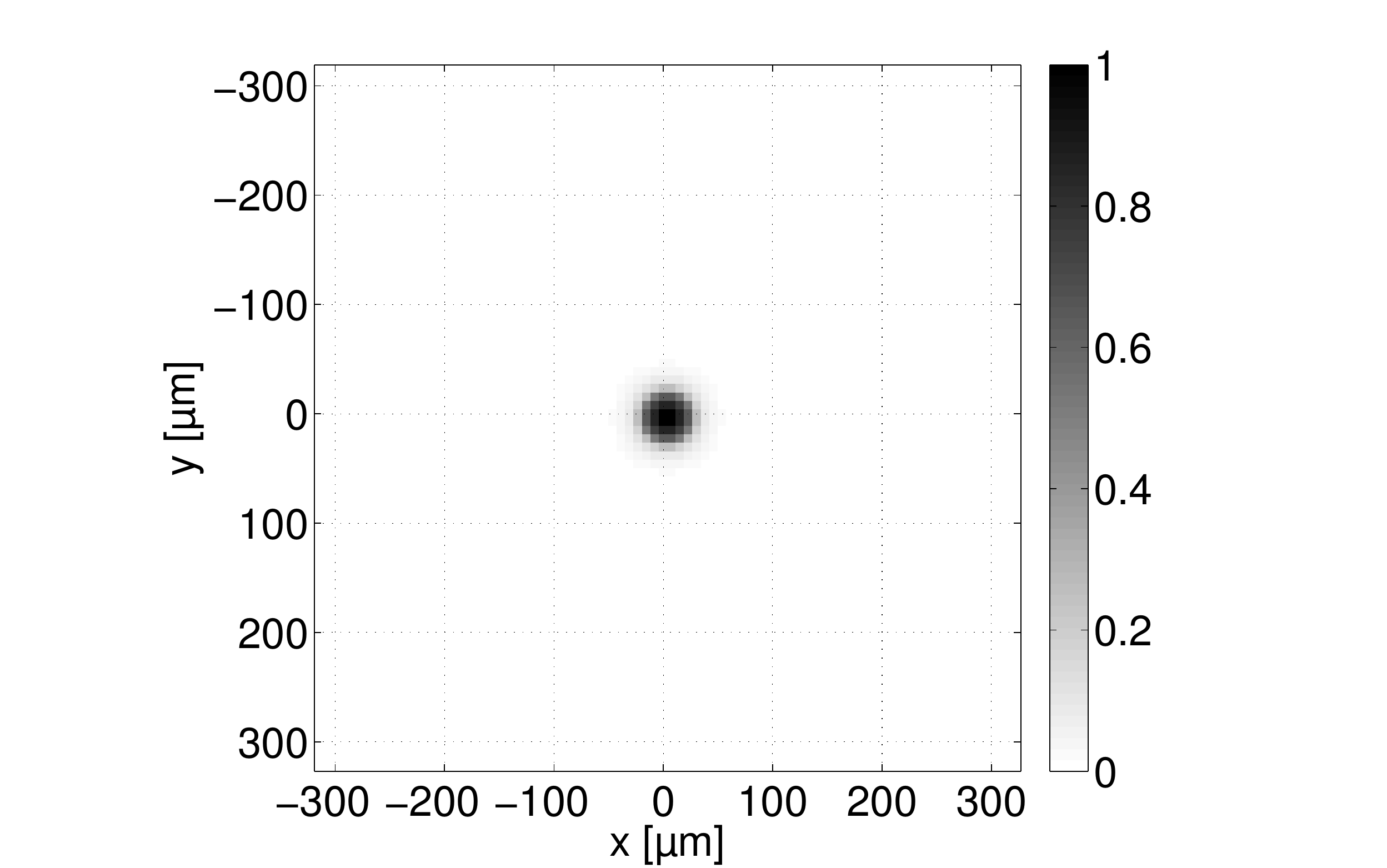}
	\end{minipage}
	\caption{\label{Image} \textbf{a}: Optical image of VUV light scattered from the collection electrode and focused onto the MCP detector surface. The analysis was performed as described in the text. \textbf{b}: Ray-tracing simulation of the measurement shown in \textbf{a}. }
\end{figure*}
Since the VUV lamp is placed in about 1.1 m distance to the CaF$_2$ entrance window, one needs to deal with very low count rates. Therefore, the MCP was set to single-photon counting mode and 800 frames, each with an exposure time of 1/16 s, were recorded resulting in a total exposure time of 50 s. 
In each frame, the positions of the single-photon peak centroids are determined. 
Using this technique, background due to internal noise of the CCD chip is reduced and the image resolution is improved. 
By plotting the single-photon peak positions, a discrete distribution, which we identify as the single channels of the MCP, becomes visible (see fig. \ref{Discrete}).
In order to obtain information on the spot size on the MCP, the single-photon peaks are replaced by filled circles that are centered at the corresponding peak centroid. 
The circles' radii are $\approx$ 40 $\mu$m and correspond to the distance between two MCP channels.
The results are shown in figs. \ref{Image}\textbf{a} and \ref{Projection}, where a gaussian FWHM of 106 $\mu$m was obtained.
Fig. \ref{Image}\textbf{b} shows a simulation of the corresponding setup, where a FWHM of about 40 $\mu$m was obtained.
The simulation underestimates the measured FWHM by a factor of about 2.5. 
This can be understood at first by reminding that the simulation represents a theoretical limit, since the ray-tracing code does not take into account shape deviations or losses and deviations due to the surface roughness of the mirrors. Secondly, the spot size on the collection electrode is not exactly 50 $\mu$m, but rather equal to the FWTM value of 80 $\mu$m, leading to an increased spot size on the MCP detector. 
In addition to that, the alignment as well as the detection occur at the limit of the MCP resolving power.

\section{Estimation of the signal-to-background ratio}
With a spot size of 106 $\mu$m and the given efficiencies (see table \ref{efficiencies}), the quality factor amounts to $\mathcal F = \epsilon/(\pi\cdot 0.053^2 \mbox{ mm}^2)=14.3\%/(8.8 \times 10^{-3} \mbox{ mm}^2)=16.3 \ (1/\mbox{mm}^2)$. $\epsilon$=14.3\% is composed of the transmission efficiency (containing the combined mirror reflectivity and the geometrical mirror acceptance) and the imaging efficiency of 23\% and 62\% respectively.
An efficiency budget is listed in table \ref{efficiencies}, where a total efficiency of 1.14 $\times10^{-5}$ is obtained. The latter gives the ratio between the number of $^{229}$Th ions that are able to leave the $^{233}$U source, \textit{i.e.} effective $^{229}$Th recoil activity, and the potentially detectable isomeric decay photons.
Presently, a $^{233}$U source is available with an effective $^{229}$Th recoil activity of 55000 1/s. 
Based on the values for efficiencies, mirror reflectivities and geometrical acceptance as listed in table \ref{efficiencies}, a signal-to-background ratio of 7000:1 is expected (see table \ref{signal-to-background}).
\begin{figure}[ht]
	\centering
	\includegraphics[width=0.45\textwidth]{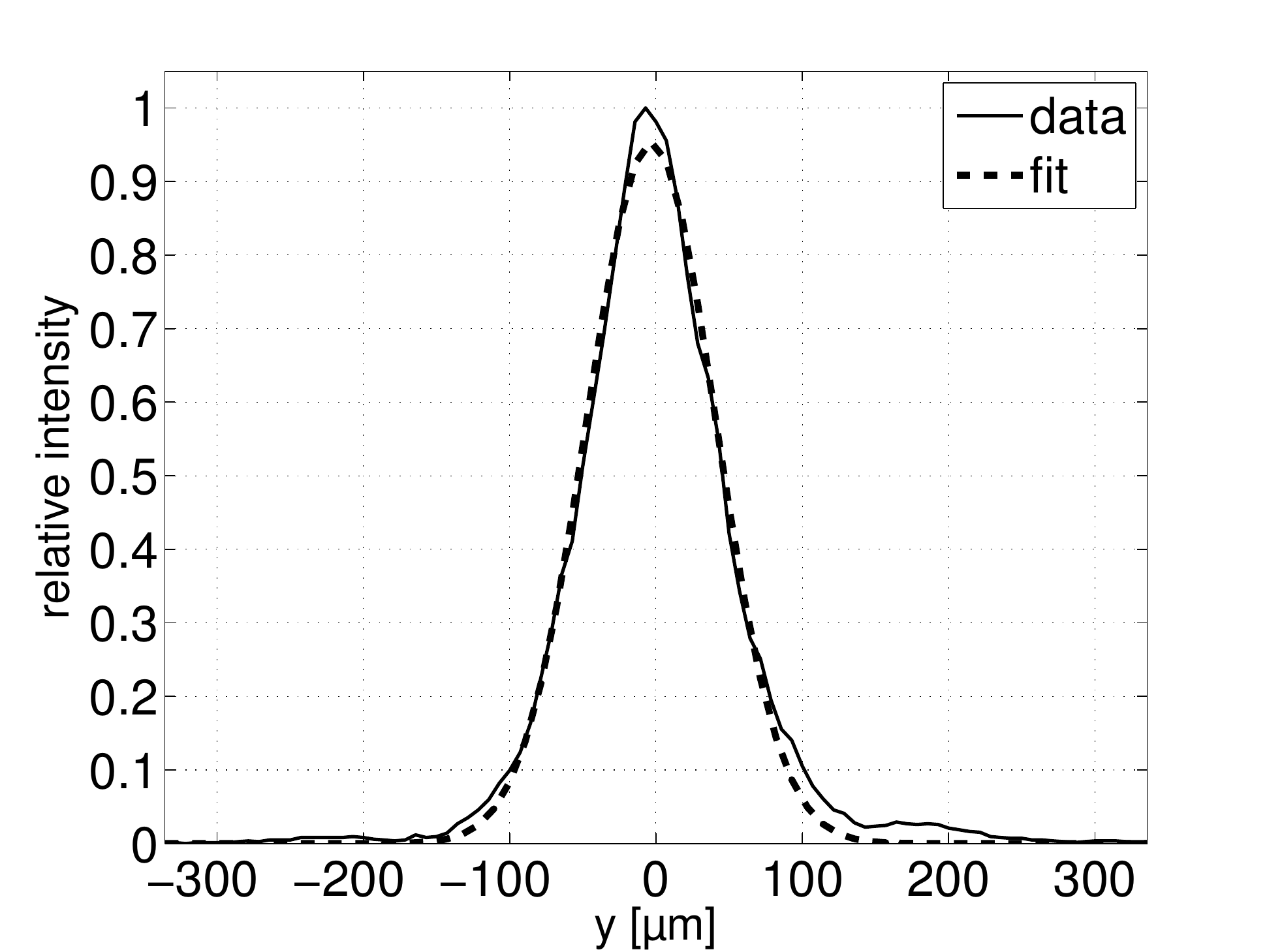}
	\caption{\label{Projection} Projection of the photon distribution shown in fig. \ref{Image}\textbf{a} onto an axis indicated in fig. \ref{Image}\textbf{a} by a vertical dashed line. The FWHM was found to be 106 $\mu$m by fitting a Gaussian function to the distribution.}
\end{figure}
\begin{table}[t]\centering
\begin{tabular}{p{.005 \textwidth} p{.26 \textwidth}  r}
			\multicolumn{3}{r}{ \textbf{efficiency [\%]} }		 \\ \hline
 \multicolumn{2}{ l}{\textbf{ion extraction part}}			&						\\[4pt]
			&	extraction of $^{229}$Th\cite{Lars}	& 10   	 \\
			&	isomer branching ratio\cite{NNDC}	&2		\\
			&	collection efficiency				&40		\\[4pt] \hline
  \multicolumn{2}{ l}{\textbf{optical setup}}			&	  			 \\ [4pt]
			& transmission effieciency				& 23			\\
			& imaging efficiency					& 62 		  	 \\ 
			& MCP efficiency at 160 nm			& 10		  	\\ [6pt] \hline \hline\\[-3pt]
\multicolumn{2}{ l}{\textbf{total efficiency:}}			& 1.14$\times10^{-3}$ \\ \hline
\end{tabular}
\caption{Efficiency values of the experimental setup. The collection efficiency is an estimate based on SIMION \cite{SIMION} simulations of the micro electrode and the triodic extraction nozzle. A detailed study of the efficiencies of the optical setup can be found in \cite{JINST}. \label{efficiencies}}
\end{table}
\begin{table}[]\centering
\begin{tabular}{l   r}
								\multicolumn{2}{r}{ \textbf{ $^{233}$U source }}	\\[4pt]\hline\\[-5pt]
effective $^{229}$Th recoil activity [1/s]			& 55000   	 \\
detected VUV photons [1/s]						&0.63		\\
count rate per area [1/(mm$^2$s)]						&71.25		\\[4pt] \hline\\[-5pt]
dark count rate per area [1/(mm$^2$s)] 		 & 0.01\\[4pt]\hline \hline\\[-3pt]
\textbf{signal-to-background ratio}						&7125:1\\\hline
\end{tabular}
\caption{Overview on the expected signal-to-background ratio for the $^{233}$U source in use. \label{signal-to-background}}
\end{table}

\section{Conclusion}
Aiming at the direct detection of the $^{229}$Th isomeric ground-state decay, a VUV optical system was developed and commissioned.
The setup was designed to meet the requirements that are set by the current knowledge on the isomer properties, promising for a successful photonic isomer detection.
To overcome the expected low count rates due to non radiative de-excitation (internal conversion, bound internal conversion or phononic coupling) the optical system provides a high light yield and efficiency.
Since the isomer's energy is not precisely known, the optical setup is optimized for a wide wavelength range above 130 nm. 
The focusing properties of the setup were extensively simulated by a ray-tracing code.\\
The optical setup is based on two annular parabolic mirrors that focus the isomeric decay radiation from $^{229}$Th ions collected on a $\O=$50 $\mu$m micro electrode onto a MCP detector. 
A test measurement of the ion collection on the micro electrode confirmed the design specifications reported in \cite{JINST}.
A special alignment procedure has been developed.
This allowed for imaging radiative processes on the micro electrode onto a spot on the MCP detector with a diameter of 106 $\mu$m (FWHM).
With the currently available $^{233}$U source and assuming a purely photonic decay of the isomer, an unsurpassed signal-to-background ratio of $\approx$ 7000:1 is expected, such that a potentially significant reduction of the photon flux (due to quenching or internal conversion) could be compensated.\\ \\
%
We acknowledge fruitful discussions with T. Udem, T. Lamour, A. Ozawa, E. Peters, T.W. H\"ansch, F. Krausz, K. Wimmer, D. Habs, P. Hilz, D. Kiefer, J. Schreiber, J. Burke, E. Haettner, M. Sondermann and the collaborators from the NuClock consortium.
This work was supported by DFG grant (Th956/3-1) and via the European Union's Horizon 2020 research and innovation programme under grant agreement No 664732 ``nuClock''.


\begin{thebibliography}{}
\bibitem{PropKrog} L.A. Kroger and C.W. Reich., Nuclear Physics A \textbf{259}, (1976) 29.
\bibitem{1990}C.W. Reich and R.G. Helmer, Phys. Rev. Lett. \textbf{64}, (1990) 271.
\bibitem{1994}C.W. Reich and R.G. Helmer, Phys. Rev. \textbf{C 49}, (1994) 1845. 
\bibitem{Beck7.6eV} B.R. Beck \textit{et al.}, Phys. Rev. Lett. \textbf{98}, (2007) 142501.
\bibitem{7.8eV} B.R. Beck \textit{et al.}, Proc. of the 12th Int. Conf. on Nucl. Reaction Mechanisms, Varenna, 2009, edited by F. Cerutti and A. Ferrari, LLNL-PROC-415170 (2009).
\bibitem{Irwin} G.M. Irwin and K.H. Kim, Phys. Rev. Lett. \textbf{79}, (1997) 6.
\bibitem{Richardson} D.S. Richardson \textit{et al.}, Phys. Rev. Lett. \textbf{80}, (1998) 15.
\bibitem{Utter} S.B. Utter \textit{et al.}, Phys. Rev. Lett. \textbf{82}, (1999) 3.
\bibitem{Shaw} R.W. Shaw \textit{et al.}, Phys. Rev. Lett. \textbf{82}, (1999) 6.
\bibitem{2001} E. Browne \textit{et al.}, Phys. Rev. C \textbf{64}, (2001) 014311.
\bibitem{2003} Barci \textit{et al.}, Phys. Rev. C \textbf{64}, (2003) 034329.
\bibitem{2003_1} T. Mitsugashira \textit{et al.}, J. Radioanal. Nucl. Ch. \textbf{255(1)}, (2003) 63.
\bibitem{2003_3}H. Kikunaga \textit{et al.}, Radiochim. Acta \textbf{93}, (2005) 507.
\bibitem{2004} I. D. Moore \textit{et al.}, Argonne National Laboratory Physics Division Report, PHY-10990-ME-2004 (2004).
\bibitem{2005} Z.O. Guimaraes and O. Helene, Phys. Rev. C \textbf{71}, (2005) 044303.
\bibitem{2005_2} Y. Kasamatsu \textit{et al.}, Research Report of Laboratory of Nuclear Science, Tohuko University, \textbf{38}, (2005) 32.
\bibitem{Zimmermann} K. Zimmermann, Ph.D. thesis, Univ. Hannover, Germany (2010).
\bibitem{Swanberg} E.L. Swanberg, Ph.D. thesis, Univ. of California, Berkeley, California, USA (2012). 
\bibitem{6+-1}X. Zhao \textit{et al.}, Phys. Rev. Lett. \textbf{109}, (2012) 160801.
\bibitem{PeikComment} E. Peik \textit{et al.}, Phys. Rev. Lett. \textbf{111}, (2013) 018901. 
\bibitem{Lifetime}E. Ruchowska \textit{et al.}, Phys. Rev. \textbf{C 73}, (2006) 044326.
\bibitem{Peik1}E. Peik and Chr. Tamm, Europhys. Lett. \textbf{61}, (2003) 181.
\bibitem{Kazakov} G.A. Kazakov \textit{et al.}, New Journal of Physics \textbf{14}, (2012) 083019.
\bibitem{Campbell} C.J. Campbell \textit{et al.}, Phys. Rev. Lett. \textbf{108}, (2012) 120802.
\bibitem{Flammbaum1} E. Litvinova \textit{et al.}, Phys. Rev. \textbf{C 79}, (2009) 064303.
\bibitem{Flammbaum2} V.V. Flambaum \textit{et al.}, Europhys. Lett. \textbf{85}, (2009) 50005.
\bibitem{Flammbaum} V.V. Flambaum, Phys. Rev. Lett. \textbf{97}, (2006) 092502.
\bibitem{Rosenbund} T. Rosenbund \textit{et al.}, Science \textbf{319}, (2008) 1808.
\bibitem{Tkalya-gamma} E.V. Tkalya, Phys. Rev. Lett. \textbf{106}, (2011) 162501.
\bibitem{Barci} V. Barci \textit{et al.}, Phys. Rev. \textbf{C 68}, (2003) 034329.
\bibitem{Haettner} Emma Haettner, Ph.D. thesis, Univ. Giessen, Germany (2011).
\bibitem{Lars} L. v.d. Wense \textit{et al.}, Eur. Phys. Journ. \textbf{A 51}, (2015) 29.
\bibitem{Karpeshin} F.F. Karpeshin \textit{et al.}, Phys. Rev. C \textbf{76}, (2007) 054313.
\bibitem{SEC-Ionization} W. F. Meggers \textit{et al.}, Natl. Bur. Stand. (U.S.), Monogr. 145 (1975). 	 
\bibitem{Micromotion} Homepage of Micromotion: \url{www.micromotion.de}.
\bibitem{Fraunhofer} R. Steinkopf \textit{et al.}, Proc. of SPIE, edited by A. Duparr\'{e} and R. Geyl, Vol. \textbf{7102} (SPIE, Bellingham, WA, 2008), 7102 0C.
\bibitem{Coating} P. Maier-Komor \textit{et al.}, Nucl. Instrum. Meth. \textbf{A 480}, (2002) 65.
\bibitem{JINST} L. v.d. Wense \textit{et al.}, J. Instrum. \textbf{8}, (2013) P03005.
\bibitem{NNDC} National Nuclear Data Center, Brookhaven National Laboratory: \url{http://www.nndc.bnl.gov}, (as of september 2015).
\bibitem{Quantar}Technical Description of the Quantar MODEL 2601B: \url{http://www.quantar.com/pdfpages/2601desc.pdf}, (2007).
\bibitem{SIMION}David A. Dahl, Int. J. Mass spectrom. \textbf{200}, (2000) 3.


\end{thebibliography}
\end{document}